\documentclass[aps,prd,showpacs,showkeys,amsmath,amssymb]{revtex4}
\usepackage{amsfonts}
\usepackage{graphicx}
\usepackage{amsmath}
\usepackage{mathrsfs}
\usepackage{dcolumn}
\usepackage{tabls}
\usepackage{slashed}
\usepackage{bm}
\usepackage{array}
\usepackage{caption2}
\allowdisplaybreaks
\begin{document}
\title{The Strong Decay Patterns of the $1^{-+}$ Exotic Hybrid Mesons}
\author{Peng-Zhi Huang}
\email{pzhuang@pku.edu.cn}
\author{Hua-Xing Chen}
\email{chx@water.pku.edu.cn}
\affiliation{Department of Physics
and State Key Laboratory of Nuclear Physics and Technology\\
Peking University, Beijing 100871, China}
\author{Shi-Lin Zhu}
\email{zhusl@pku.edu.cn}
\affiliation{Department of Physics
and State Key Laboratory of Nuclear Physics and Technology\\
and Center of High Energy Physics, Peking University, Beijing
100871, China }

\begin{abstract}

We calculate the coupling constants of the decay modes
$1^{-+}\rightarrow\rho\pi, f_1\pi, b_1\pi, \eta\pi, \eta'\pi,
a_1\pi, f_1\eta$ within the framework of the light-cone QCD sum
rule. Then we calculate the partial width of these decay channels,
which differ greatly from the existing calculations using
phenomenological models. For the isovector $1^{-+}$ state, the
dominant decay modes are $\rho\pi, f_1\pi$. For its isoscalar
partner, its dominant decay mode is $a_1\pi$. We also discuss the
possible search of the $1^{-+}$ state at BESIII, for example
through the decay chains $J/\psi (\psi')\to \pi_1 +\gamma$ or
$J/\psi (\psi')\to \pi_1 +\rho$ where $\pi_1$ can be reconstructed
through the decay modes $\pi_1\to \rho\pi\to \pi^+\pi^-\pi^0$ or
$\pi_1\to f_1(1285)\pi^0$. Hopefully the present work will be
helpful to the experimental establishment of the $1^{-+}$ hybrid
meson.

\end{abstract}
\maketitle
\pagenumbering{arabic}

\section{Introduction}\label{introduction}

Quark model has been proved to be very successful in the
classification of hadrons and calculation of hadron spectrum and
their other properties. Yet Quantum Chromodynamics (QCD), which is
widely accepted as the fundamental theory of the strong
interaction, does not prohibit the existence of those hadron
states which can not be accommodated in the conventional quark
model. These non-conventional hadrons include multi-quark states
($qq\bar{q}\bar{q}$, $qqqq\bar{q}$, $\cdots$), glueballs ($gg$,
$ggg$, $\cdots$), and hybrids ($q\bar{q}g$). Searching for
non-conventional hadrons experimentally and studies on their
properties have attracted much interests over the past few
decades. With these efforts one aims to explore the
nonperturbative aspects of QCD in the low energy sector.

Some non-conventional hadrons are totally ``exotic'', namely their
$J^{PC}$ quantum numbers are excluded by the conventional quark
model. For example, a conventional $q\bar q$ meson possesses
parity $P=(-1)^{L+1}$ and $C$-parity $C=(-1)^{L+S}$, where $L$ and
$S$ are the orbital angular momentum and spin of its componential
quark and antiquark, respectively. Apparently,
$J^{PC}=0^{--},0^{+-},1^{-+},2^{+-}$ etc. are impossible for
conventional mesons. Hadrons with these $J^{PC}$ quantum numbers
are exotic states, which are widely studied since they do not mix
with conventional hadrons.

Recently, COMPASS collaboration observed a resonance with exotic
quantum numbers $J^{PC}=1^{-+}$ at $(1660\pm10^{+0}_{-64})\
\text{MeV}/c^2$ with a width of $(269\pm21^{+42}_{-64})\
\text{MeV}/c^2$ \cite{compass}. In literature, three isovector
$J^{PC}=1^{-+}$ exotic mesons, namely $\pi_1(1400)$,
$\pi_1(1600)$, and $\pi_1(2015)$, have been reported. Several
groups observed $\pi_1(1400)$ in the $\eta\pi^-$ system
\cite{1400}. $\pi_1(1600)$ was first reported through a combined
study of the $\eta'\pi^-$, $f_1\pi^-$, and $\rho^0\pi^-$ channels
by VES in Ref. \cite{1600VES1} and later in the $b_1\pi$ final
state in Ref. \cite{1600VES2}. E852 collaboration observed
$\pi_1(1600)$ by carrying out the partial wave analysis in the
$\pi^+\pi^-\pi^-$ final state \cite{1600E8521} and examining the
$\eta'\pi^-$ final state in the reaction $\pi^-p\rightarrow
p\eta'\pi^-$ \cite{1600E8522}. They also found it in the final
states $f_1\pi$ \cite{1600E8523} and $b_1\pi$ \cite{1600E8524}.
The Crystal Barrel collaboration analyzed the reaction
$\bar{p}p\rightarrow \omega\pi^+\pi^-\pi^0$ and observed the
$\pi_1(1600)$ in the $b_1\pi$ channel in Ref. \cite{1600Crystal}.
E852 also reported another $1^{-+}$ meson $\pi_1(2015)$
\cite{1600E8523, 1600E8524}.

The $1^{-+}$ hybrids have been studied in a few different
theoretical schemes. The mass of the lowest-lying $1^{-+}$ hybrid
meson was predicted to be around $1.9\ \text{GeV}$ in the flux
tube model \cite{fluxtubemass}. Llanes-Estrada and Cotanch
predicted the hybrid mass to be above $2.0\ \text{GeV}$ utilizing
a QCD inspired Coulomb gauge Hamiltonian \cite{coulombhamilton}.
H.-C. Kim and Y. Kim investigated the $1^{-+}$ hybrid within the
framework of an AdS/QCD model \cite{AdSQCD}. Early studies with
the leading order QCD sum rule estimated the mass of the $1^{-+}$
hybrid to be $(1.6\sim2.1)\ \text{GeV}$ \cite{QSRmass1}. The
inclusion of the radiative corrections changed this estimate to
$1.26\pm0.15\ \text{GeV}$ \cite{QSRmass2}, $(1.6\sim1.7)\
\text{GeV}$ \cite{QSRmass3}, and $1.81(6)\ \text{GeV}$
\cite{QSRmass4}. The Lattice QCD prediction for the mass of
$1^{-+}$ falls into a wide range of $(1.5\sim2.2)\ \text{GeV}$
\cite{latticemass}.

The $1^{-+}$ hybrids were shown to exist as narrow resonant states
in the large $N_c$ limit of QCD \cite{largeNcmass}. The coupling
of a neutral hybrid $\{1, 3, 5\dots\}^{-+}$ to two neutral
(hybrid) mesons with the same $J^{PC}$ and $J=0$ were studied in
the large $N_c$ limit of QCD in Ref. \cite{largeNcdecay}. Cook and
Fiebig presented a decay width calculation on lattice for the
channel $1^{-+}\rightarrow a_1\pi$ in Ref. \cite{latticedecay1}.
The partial widths of the ground $1^{-+}$ hybrid to $b_1\pi$ and
$f_1\pi$ were predicted to be $400(120)\ \text{MeV}$ and $90(60)\
\text{MeV}$, respectively in Ref. \cite{latticedecay2}. The strong
decay properties were explored in Ref. \cite{IKP} (IKP) within the
same framework. Page, Swanson, and Szczepaniak \cite{PSS} (PSS)
extended the original IKP flux tube model and studied the strong
decays of hybrid mesons with different $J^{PC}$ quantum numbers,
including those of the $1^{-+}$ hybrids. Burns and Close
\cite{fluxtubelattice} compared the flux tube model and Lattice
QCD for the $S$-wave decay of the $1^{-+}$ hybrid and found
excellent agreement.

Some decay modes of the isoscalar and isovector $1^{-+}$ hybrids
were studied using the three-point function sum rule
\cite{govaerts}. In Ref. \cite{pascual}, the decay widths of the
$1^{-+}$ hybrid were calculated using a three-point function sum
rule evaluated at the symmetric Euclidean point. The mass of the
strange hybrid was also studied in this paper within a light quark
expansion formalism. The partial width of the channel
$1^{-+}\rightarrow\rho\pi$ was predicted to be rather broad by
using the three-point function at the symmetric point
\cite{pascual, narisonbook}. Zhu reexamined the decay channels
$1^{-+}\rightarrow\rho\pi,f_1\pi$ by using the light-cone QCD sum
rules and reduced the partial width of $1^{-+}\rightarrow\rho\pi$
significantly \cite{zhu}.

In this work, we employ Light-cone QCD sum rule (LCQSR)
\cite{light-cone} to calculate the various coupling constants of
the decay modes $1^{-+}\rightarrow\rho\pi, f_1\pi, b_1\pi,
\eta\pi, \eta'\pi, a_1\pi, f_1\eta$. The LCQSR is different from
the traditional Shifman-Vainshtein-Zakharov (SVZ) sum rule
\cite{svz}, where the short-distance OPE expansion is made. The
OPE expansion of the LCQSR is performed near the light cone. With
the extracted coupling constants, we calculate the partial widths
of these modes and compare them with the existing results obtained
using other approaches. We also suggest the possible search for
the $1^{-+}$ hybrid mesons at BESIII.

The paper is organized as follows. We illustrate the formalism and
derive the LCQSR for the $\pi_1\rightarrow\rho\pi$ coupling
constant in Sec. \ref{rhocoupling}. We present the sum rules for
the $\pi_1\rightarrow f_1\pi,b_1\pi$ and $\pi_1\rightarrow
\eta\pi,\eta'\pi$ coupling constants in Sec. \ref{f1b1coupling}
and \ref{gammacoupling} respectively. The decay widths of the
isovector and isoscalar $1^{-+}$ states are presented in Sec.
\ref{decaywidth} and \ref{isoscalar} respectively. We discuss the
possible search of these exotic states at BESIII in Sec.
\ref{bes}. The last section is a short summary. The light cone
distribution amplitudes of the pion which are employed in the
present calculation are collected in the appendix. Readers who are
not interested in the technical details may skip Secs.
\ref{rhocoupling}-\ref{gammacoupling} and go to last three
sections directly.

\section{Sum rules for the $\pi_1\rightarrow\rho\pi$ coupling constant}\label{rhocoupling}

We denote the isovector and the isoscalar $J^{PC}=1^{-+}$ hybrid meson by $\pi_1$ and $\tilde{\pi}_1$, respectively.
The adopted interpolating current for $\pi_1$ reads
\begin{eqnarray}
J^{\pi_1}_\mu(x)=\frac{1}{\sqrt{2}}\left[\bar{u}(x)\frac{\lambda^a}{2}g_sG^a_{\mu\nu}(x)\gamma^\nu
u(x) -\bar{d}(x)\frac{\lambda^a}{2}g_sG^a_{\mu\nu}(x)\gamma^\nu
d(x)\right]\,.
\end{eqnarray}
The overlapping amplitude $\tilde{f}_{\pi_1}$ between the above interpolating current and $\pi_1$ is defined as
\begin{eqnarray}
\langle 0|J^{\pi_1}_\mu(0)|\pi_1(p,\lambda)\rangle=\tilde{f}_{\pi_1}\eta^\lambda_\mu\,,
\end{eqnarray}
where $\eta_\mu$ is the polarization vector of $\pi_1$.

We consider the following correlation functions in our calculation:
\begin{eqnarray}
\Pi_\rho(k^2,p^2)=i\int d^4 xe^{ik\cdot x}\langle \pi(q)|T\{J^{\rho}_\alpha(x)J^{\pi_1\dagger}_\beta(0)\}|0\rangle
&=&i\varepsilon_{\alpha\beta\gamma\delta}k^\gamma q^\delta G_\rho(k^2,p^2)+\cdots\,,
\end{eqnarray}
where $p=k+q$. The interpolating current for the $\rho$ meson is
$J^\rho_\mu(x)=\bar{d}(x)\gamma_\mu u(x)$ and we have $\langle
0|J^\rho_\mu(0)|\rho(k,\lambda)\rangle=f_\rho
m_\rho\epsilon^\lambda_\mu$ where $\epsilon_\mu$ is the
polarization vector of the $\rho$ meson. The decay amplitude of
the channel $\pi_1\rightarrow\rho\pi$ can be written as
\begin{eqnarray}
\mathcal{M}(\pi_1\rightarrow\rho\pi)
=\varepsilon_{\alpha\beta\gamma\delta}\epsilon^{*\alpha}\eta^\beta k^\gamma q^\delta g_\rho\,,
\end{eqnarray}
where $k$ and $q$ are the momentum of the final $\rho$ and $\pi$, respectively.
When $k^2,p^2\ll0$, $G_\rho(k^2,p^2)$ can be calculated by operator product expansion(OPE) near the light-cone $x^2=0$,
with the $\pi$ light-cone wave functions as its input.
Furthermore, $G_\rho(k^2,p^2)$ can be related to $g_\rho$ by the dispersion relation
\begin{eqnarray}\label{dispersionrelation}
G_\rho(k^2,p^2)
=\int_0^\infty ds_1\int_0^\infty ds_2 \frac{\rho(s_1,s_2)}{(s_1-k^2-i\epsilon)(s_2-p^2-i\epsilon)}
+\int_0^\infty ds_1\frac{\rho_1(s_1)}{s_1-k^2-i\epsilon}
+\int_0^\infty ds_2\frac{\rho_2(s_2)}{s_2-p^2-i\epsilon}
+\cdots\,,
\end{eqnarray}
where
\begin{eqnarray}
\rho(s_1,s_2)=f_\rho \tilde{f}_{\pi_1}m_\rho g_\rho\delta(s_1-m_\rho^2)\delta(s_2-m_{\pi_1}^2)+\cdots\,.
\end{eqnarray}

\begin{figure}[!htb]
\centering
\begin{minipage}[t]{0.35\linewidth}
\includegraphics[width=1.75in]{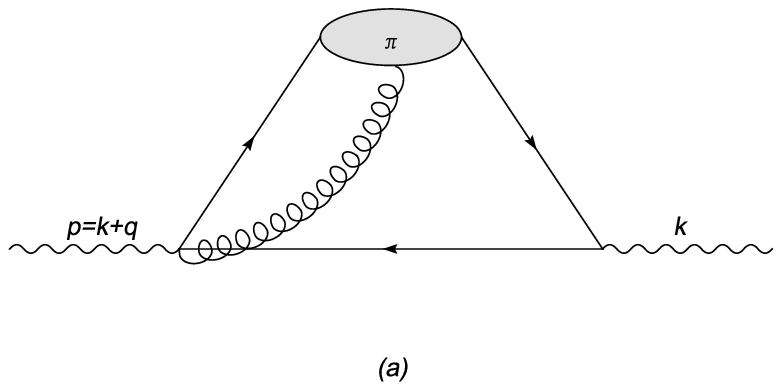}
\end{minipage}%
\begin{minipage}[t]{0.35\linewidth}
\includegraphics[width=1.75in]{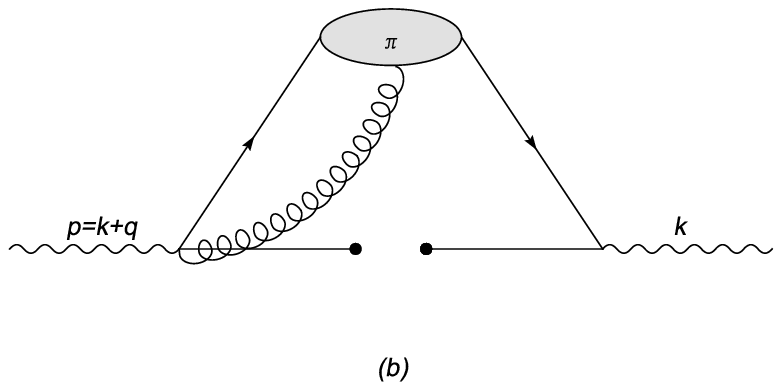}
\end{minipage}
\end{figure}

\begin{figure}[!htb]
\centering
\begin{minipage}[t]{0.3\linewidth}
\includegraphics[width=1.75in]{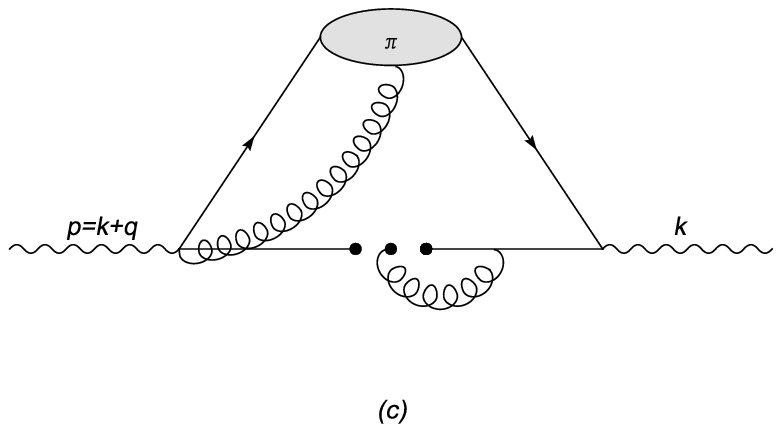}
\end{minipage}%
\begin{minipage}[t]{0.3\linewidth}
\includegraphics[width=1.75in]{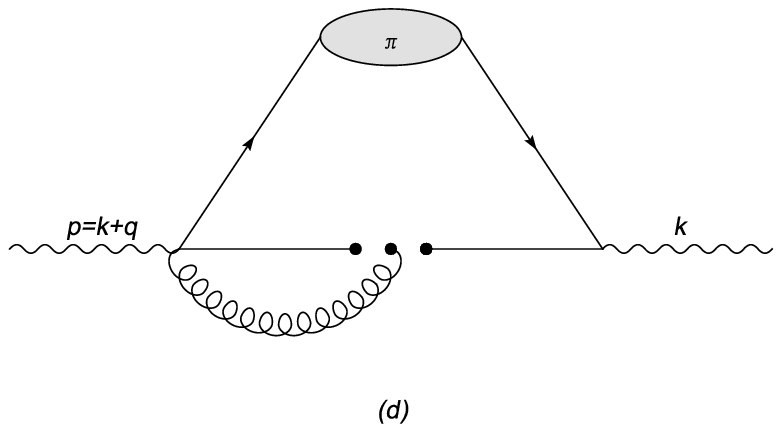}
\end{minipage}%
\begin{minipage}[t]{0.3\linewidth}
\includegraphics[width=1.75in]{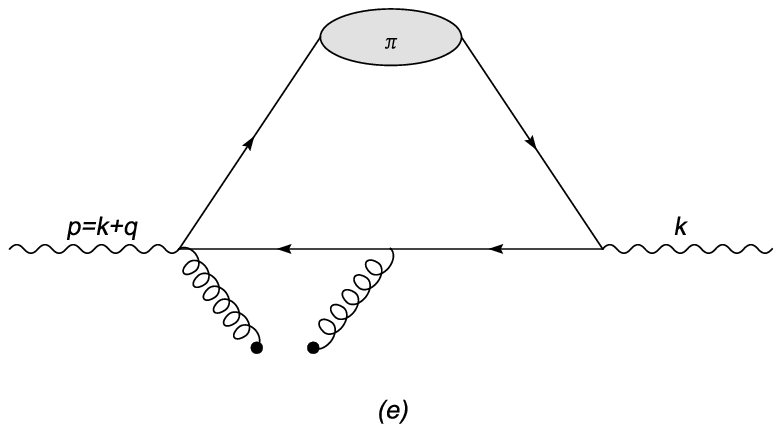}
\end{minipage}
\caption{The Feynman diagrams for $\Pi_\rho(k^2,p^2)$.}
\label{fig:feynDiag}
\end{figure}

After invoking the double Borel transformation $\mathcal{B}_{k^2}^{M_1^2}\mathcal{B}_{p^2}^{M_2^2}$,
we extract the double dispersion relation part of Eq. (\ref{dispersionrelation}):
\begin{eqnarray}\label{rhopresumrule}
&&f_\rho \tilde{f}_{\pi_1}m_\rho g_\rho e^{\bar{u}_0m_\rho^2/M^2+u_0m_{\pi_1}^2/M^2}+\cdots\nonumber\\
&&=e^{u_0\bar{u}_0m_\pi^2/M^2}
\biggl\lbrace\frac{f_\pi}
{\sqrt{2}}(\mathcal{A}_\perp^{[\alpha_2]}-\mathcal{A}_\perp^{[\alpha_1]}-\mathcal{V}_\parallel^{[\alpha_1]}
-\mathcal{V}_\parallel^{[\alpha_2]}+\mathcal{V}_\perp^{[\alpha_1]}+\mathcal{V}_\perp^{[\alpha_2]})m_\pi^2M^2
-\frac{f_\pi}{36\sqrt{2}}[\phi_\pi(u_0)+\phi_\pi(\bar{u}_0)]\langle g_s^2G^2\rangle\biggl\rbrace\nonumber\\
&&\approx\frac{f_\pi}
{\sqrt{2}}(\mathcal{A}_\perp^{[\alpha_2]}-\mathcal{A}_\perp^{[\alpha_1]}-\mathcal{V}_\parallel^{[\alpha_1]}
-\mathcal{V}_\parallel^{[\alpha_2]}+\mathcal{V}_\perp^{[\alpha_1]}+\mathcal{V}_\perp^{[\alpha_2]})m_\pi^2M^2
-\frac{f_\pi}{36\sqrt{2}}[\phi_\pi(u_0)+\phi_\pi(\bar{u}_0)]\langle g_s^2G^2\rangle\,,
\end{eqnarray}
where
\begin{eqnarray}
 u_0=\frac{M_1^2}{M_1^2+M_2^2},\ \ \ \ \ \ M^2=\frac{M_1^2M_2^2}{M_1^2+M_2^2}\,,
\end{eqnarray}
and $\bar{x}\equiv1-x$.
Hereafter we ignore the factor $e^{u_0\bar{u}_0m_\pi^2/M^2}$
because $m_\pi^2/M^2<0.01$ in our calculations.
The definitions of $\mathcal{F}^{[\alpha_i]}$s are
\begin{eqnarray}
\mathcal{F}^{[\alpha_1]}&\equiv&\int_0^{\bar{u}_0}\mathcal{F}(\alpha_1,u_0,\bar{u}_0-\alpha_1)\,d\alpha_1\,,\nonumber\\
\mathcal{F}^{[\alpha_2]}&\equiv&\int_0^{\bar{u}_0}\mathcal{F}(u_0,\alpha_2,\bar{u}_0-\alpha_2)\,d\alpha_2\,,\nonumber\\
\mathcal{F}^{[\alpha_1,u]}&\equiv&\int_0^{\bar{u}_0}\int_0^{u}\mathcal{F}(\alpha_1,\alpha_2',\bar{\alpha}_1-\alpha_2')\,d\alpha_2'd\alpha_1\,,\nonumber\\
\mathcal{F}^{[\alpha_2,u]}&\equiv&\int_0^{\bar{u}_0}\int_0^{u}\mathcal{F}(\alpha_1',\alpha_2,\bar{\alpha}_2-\alpha_1')\,d\alpha_1'd\alpha_2\,.
\end{eqnarray}
Here the Borel transformation is defined as
\begin{eqnarray}
\mathcal{B}_{k^2}^{M_1^2}[f(k^2)]
=\lim_{n\rightarrow\infty}\frac{(-k^2)^{n+1}}{n!}\left(\frac{d}{dk^2}\right)^nf(k^2)\left|_{k^2=-nM_1^2}\right.\,.
\end{eqnarray}
The quark propagator used in the OPE of $G_\rho(k^2,p^2)$ is
\begin{eqnarray}\nonumber
iS(x)
=\frac{i\slashed{x}}{2\pi^2x^4}
+\frac{i}{32\pi^2}\frac{\lambda^n}{2}g_sG^n_{\mu\nu}\frac{1}{x^2}(\sigma^{\mu\nu}\slashed{x}+\slashed{x}\sigma^{\mu\nu})
-\frac{\langle\bar qq\rangle}{12}
-\frac{m_0^2\langle\bar qq\rangle x^2}{192}
+\cdots\,.
\end{eqnarray}
We present the Feynman diagrams corresponding to the quark-level calculation of $\Pi_\rho(k^2,p^2)$ in Fig. \ref{fig:feynDiag}.

The spectral density $\rho(s_1,s_2)$ can be derived from the
dispersion realtion Eq. (\ref{dispersionrelation}) after two
continuous double Borel transformation of $G_\rho(k^2,p^2)$:
\begin{eqnarray}
\rho(s_1,s_2)=\mathcal{B}_{\sigma_1}^{\frac{1}{s_1}}\mathcal{B}_{\sigma_2}^{\frac{1}{s_2}}
\mathcal{B}_{k^2}^{\frac{1}{\sigma_1}}\mathcal{B}_{p^2}^{\frac{1}{\sigma_2}}G_\rho(k^2,p^2)\,.
\end{eqnarray}
According to quark-hadron duality,
we can subtract the contribution of the excited states and the continuum from Eq. (\ref{rhopresumrule}) and arrive at
\begin{eqnarray}
f_\rho \tilde{f}_{\pi_1}m_\rho g_\rho e^{\bar{u}_0m_\rho^2/M^2+u_0m_{\pi_1}^2/M^2}
=\int_0^{s_{01}} ds_1\int_0^{s_{02}} ds_2\;e^{-s_1\sigma_1}e^{-s_2\sigma_2}\;
\mathcal{B}_{\sigma_1}^{\frac{1}{s_1}}\mathcal{B}_{\sigma_2}^{\frac{1}{s_2}}
\mathcal{B}_{k^2}^{\frac{1}{\sigma_1}}\mathcal{B}_{p^2}^{\frac{1}{\sigma_2}}G_\rho(k^2,p^2)\,,
\end{eqnarray}
where $s_{01}$ and $s_{02}$ are the continuum thresholds of the mass rules of the $\rho$ meson and the hybrid state $\pi_1$, respectively.

The large mass difference between the $\pi_1$ hybrid and the $\rho$
meson inclines us to work at an asymmetric point of the Borel
parameter $M_1^2$ and $M_2^2$, leading to a sophisticated
subtraction of the continuum contribution \cite{asymmetricpoint}.
The terms of
$\mathcal{B}_{k^2}^{\frac{1}{\sigma_1}}\mathcal{B}_{p^2}^{\frac{1}{\sigma_2}}G_\rho(k^2,p^2)$
have general form
$cu_0^m(M^2)^n=c\sigma_2^m/(\sigma_1+\sigma_2)^{m+n}$. Here we
assume $m,n>0$ to illustrate the procedure of the continuum
subtraction.
\begin{eqnarray}
&&\int_0^{s_{01}} ds_1\int_0^{s_{02}} ds_2\;e^{-s_1\sigma_1}e^{-s_2\sigma_2}\;
\mathcal{B}_{\sigma_1}^{\frac{1}{s_1}}\mathcal{B}_{\sigma_2}^{\frac{1}{s_2}}
\frac{\sigma_2^m}{(\sigma_1+\sigma_2)^{m+n}}\nonumber\\
&&=\int_0^{s_{01}} ds_1\int_0^{s_{02}} ds_2\;e^{-s_1\sigma_1}e^{-s_2\sigma_2}\;
\frac{1}{\Gamma(m+n)}\left[-\frac{\partial\delta(s_1-s_2)}{\partial s_1}\right]^ms_1^{m+n-1}\nonumber\\
&&=2\int_0^{s_{01}} ds_+\int_{-s_+}^{s_+} ds_-\;e^{-s_+M^2}e^{s_-M_-^2}\;
\frac{(s_+-s_-)^{m+n-1}}{2^m\Gamma(m+n)}\left(\frac{\partial}{\partial s_-}\right)^m\delta(2s_-)\nonumber\\
&&=\frac{M^{2n}}{2^m}\sum^m_{i=0}\frac{m!}{i!(m-i)!}(2u_0-1)^if_{n-1+i}(\frac{s_{01}}{M^2})\,,
\end{eqnarray}
where $s_+=(s_1+s_2)/2$, $s_-=(s_2-s_1)/2$, $1/M_-^2=1/M_1^2-1/M_2^2$ and we assume $s_{01}<s_{02}$\,.
$f_n(x)$ is the subtraction function defined as
\begin{eqnarray}
f_n(x)=1-e^{-x}\sum^n_{i=0}\frac{x^i}{i!}\,.
\end{eqnarray}

The $1^{-+}$ hybrid has not been firmly established
experimentally. Both theoretical predictions and experimental
measurements suggest that the mass of $\pi_1$ falls within the range
$1.6\sim2.0\ \text{GeV}$. In this work the mass of $\pi_1$ is taken
to be $m_{\pi_1}=1.6$, $1.8$ and $2.0\ \text{GeV}$. We adopt
$\tilde{f}_{\pi_1}=0.15\ \text{GeV}^4$ in our numerical analysis
\cite{QSRmass3}. The $\pi$ decay constant $f_\pi=131\ \text{MeV}$.
The mass and the decay constant of the $\rho$ meson are
$m_\rho=0.77\ \text{GeV}$ and $f_\rho=0.216\ \text{GeV}$.
$\mu_\pi\equiv m_\pi^2/(m_u+m_d)=(1.573\pm 0.174)\ \text{GeV}$ is
given in Ref. \cite{pilcda}.

The parameters appear in the $\pi$ distribution amplitudes are listed below \cite{pilcda}.
We use the values at the scale $\mu=1\ \text{GeV}$ in our calculation.
\begin{center}
\setlength\extrarowheight{8pt}
\begin{tabular}{ccccccccccc}
\hline
  $a_2$\ \  &$\eta_3$\ \  &$\omega_3$\ \   &$\eta_4$\ \  &$\omega_4$\ \  &$h_{00}$\ \  &$v_{00}$\ \  &$a_{10}$\ \  &$v_{10}$\ \  &$h_{01}$\ \  &$h_{10}$\\
  $0.25$\ \ &$0.015$\ \   &$-1.5$\ \       &$10$\ \      &$0.2$\ \       &$-3.33$\ \   &$-3.33$\ \   &$5.14$\ \    &$5.25$\ \    &$3.46$\ \    &$7.03$  \\
\hline
\end{tabular}
\end{center}

It is reasonable to let $M_1^2=\beta m_\rho$ and $M_2^2=\beta m_{\pi_1}$,
where $\beta$ is a dimensionless scale parameter.
Then we have $u_0=m_\rho^2/(m_\rho^2+m_{\pi_1}^2)$ and $M^2=\beta m_\rho^2m_{\pi_1}^2/(m_\rho^2+m_{\pi_1}^2)$.

\begin{figure}[!htb]
\captionstyle{flushleft}
\includegraphics[width=3in]{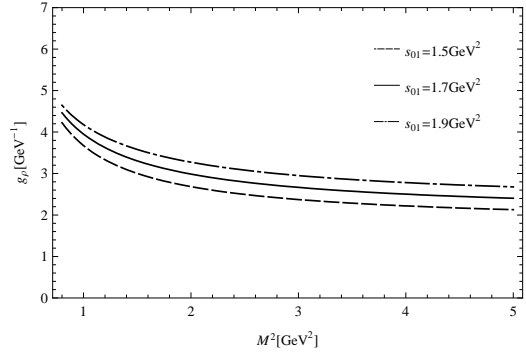}
\caption{The sum rule for $g_\rho$ with $m_{\pi_1}=1.6\ \text{GeV}$, $2.3<M^2<2.7\ \text{GeV}^2$, and $s_{01}=1.5,1.7,1.9\ \text{GeV}^2$.}
\label{fig:CCRhoPim16}
\end{figure}

From the requirement of the convergence of the OPE and the
requirement that the pole contribution is larger than $40\%$, we
get the working interval of the Borel parameter $M^2$. The
resulting sum rule is plotted with $s_{01}=1.5,1.7,1.9\
\text{GeV}^2$ in Fig. \ref{fig:CCRhoPim16} in the case of
$m_{\pi_1}=1.6\ \text{GeV}$. The sum rules for $m_{\pi_1}=1.8,2.0\
\text{GeV}$ are similar. The numerical values of $g_\rho$ are
presented here with their variations determined by the working
interval of the Borel parameter $2.3<M^2<2.7\ \text{GeV}^2$ and
the range of the threshold $1.5<s_{01}<1.9\ \text{GeV}^2$.
\begin{center}
\setlength\extrarowheight{8pt}
\begin{tabular}{cccc}
\hline
$m_{\pi_1}[\text{GeV}]$         & $1.6$                    & $1.8$                    & $2.0$                     \\
$g_\rho[\text{GeV}^{-1}]$     & $\quad2.5\sim3.2\quad$   & $\quad2.6\sim3.3\quad$   & $\quad2.6\sim3.4\quad$              \\
\hline
\end{tabular}
\end{center}

\section{Sum rules for the $\pi_1\rightarrow f_1\pi,b_1\pi$ coupling constants}\label{f1b1coupling}

Using the method outlined above we can further derive the sum
rules for the $\pi_1\rightarrow f_1\pi,b_1\pi$ coupling constants.
The effective Lagrangians of the process $\pi_1\rightarrow f_1\pi$
can be written as
\begin{eqnarray}
\mathcal{L}_{f_1}
&=& g^1_{f_1}    \vec\pi_{1\alpha} \times \vec f_{1\beta} \cdot \vec\pi g^{\alpha\beta} \nonumber\\
&&+ g^{21}_{f_1} \partial^\beta \vec\pi_{1\alpha} \times \partial^\alpha \vec f_{1\beta} \cdot \vec \pi
  + g^{22}_{f_1} \partial^\beta \vec\pi_{1\alpha} \times \vec f_{1\beta} \cdot \partial^\alpha \vec \pi \nonumber\\
&&+ g^{23}_{f_1} \vec\pi_{1\alpha} \times \partial^\alpha \vec f_{1\beta} \cdot \partial^\beta \vec \pi
  + g^{24}_{f_1} \vec\pi_{1\alpha} \times \vec f_{1\beta} \cdot \partial^\alpha\partial^\beta \vec \pi \,.
\end{eqnarray}
Then the decay amplitude for this process is
\begin{eqnarray}
\mathcal {M}(\pi_1\rightarrow f_1+\pi)
=ig_{f_1}^1(\eta\cdot\epsilon^*)+ig_{f_1}^2(\eta\cdot k)(\epsilon^*\cdot p)\,,
\end{eqnarray}
where $g_{f_1}^2=g_{f_1}^{21}-g_{f_1}^{22}+g_{f_1}^{23}-g_{f_1}^{24}$,
$\epsilon^*$ is the polarization vector of the final $f_1$ meson,
$k$ and $p$ are the momentum of the $f_1$ and the $\pi_1$ meson, respectively.
The interpolating current for the $f_1$ meson is the axial-vector current
\begin{eqnarray}\label{f1current}
J^{f_1}_\mu(x)=\frac{1}{\sqrt{2}}\left[\bar{u}(x)\gamma_\mu\gamma_5 u(x)
+\bar{d}(x)\gamma_\mu\gamma_5 d(x)\right]\,.
\end{eqnarray}
The $f_1$ meson couples to the above current through
$\langle 0|J^{f_1}_\mu(0)|f_1(k,\lambda)\rangle=f_{f_1}m_{f_1}\epsilon^\lambda_\mu$.
We consider the following correlation functions:
\begin{eqnarray}
\Pi_{f_1}(k^2,p^2)&=&i\int d^4 xe^{ik\cdot x}\langle \pi(q)|T\{J^{f_1}_\alpha(x)J^{\pi_1\dagger}_\beta(0)\}|0\rangle\nonumber\\
&=& -\big(g_{\alpha\nu}-{p_\alpha p_\nu \over m_{\pi_1}^2} \big)\big(g_{\beta\mu}-{k_\beta k_\mu \over m_{f_1}^2}\big)
    {\tilde{f}_{\pi_1} f_{f_1} m_{f_1} \over (m_{\pi_1}^2-p^2) (m_{f_1}^2-k^2)} \nonumber\\
&&  \times \left[g^1_{f_1} g_{\mu\nu}+g^{21}_{f_1}k_\mu p_\nu
    +g^{22}_{f_1}q_\mu p_\nu+g^{23}_{f_1}k_\mu q_\nu+g^{24}_{f_1} q_\mu q_\nu \right]+\cdots\nonumber\\
&=& -\big(g_{\alpha\nu}-{p_\alpha p_\nu \over m_{\pi_1}^2} \big)\big(g_{\beta\mu}-{k_\beta k_\mu \over m_{f_1}^2}\big)
    {\tilde{f}_{\pi_1} f_{f_1} m_{f_1} \over (m_{\pi_1}^2-p^2) (m_{f_1}^2-k^2)} \nonumber\\
&&  \times \left[g^1_{f_1} g_{\mu\nu}+g^{21}_{f_1}k_\mu p_\nu+ g^{22}_{f_1}(p_\mu-k_\mu) p_\nu
    +g^{23}_{f_1}k_\mu (p_\nu-k_\nu)+g^{24}_{f_1} (p_\mu-k_\mu) (p_\nu-k_\nu) \right]+\cdots\nonumber\\
&=& -{\tilde{f}_{\pi_1} f_{f_1} m_{f_1} \over (m_{\pi_1}^2-p^2) (m_{f_1}^2-k^2)} \left[g^1_{f_1} g_{\alpha\beta}
     +(g^{21}_{f_1}-g^{22}_{f_1}+g^{23}_{f_1}-g^{24}_{f_1})p_\alpha k_\beta\right]+\cdots\nonumber\\
&=& -{\tilde{f}_{\pi_1} f_{f_1} m_{f_1} \over (m_{\pi_1}^2-p^2) (m_{f_1}^2-k^2)} \left[g^1_{f_1} g_{\alpha\beta}+g^2_{f_1}p_\alpha k_\beta\right]+\cdots\,.
\end{eqnarray}
The axial-vector current (\ref{f1current}) also
couples to $I=0$ pseudoscalar meson $\eta/\eta'$. However we can
differentiate the contribution of the $P$-wave channel
$\pi_1\rightarrow\eta/\eta'\pi$ to the correlation function from
that of $\pi_1\rightarrow f_1\pi$ due to their different Lorentz
structures. In a similar way to Section. {\ref{rhocoupling}}, we
obtain the following sum rules for $g_{f_1}^1$ and $g_{f_1}^2$
before the continuum subtraction:
\begin{eqnarray}\label{f1presumrule1}
&&f_{f_1}\tilde{f}_{\pi_1}m_{f_1}g_{f_1}^1 e^{\bar{u}_0m_{f_1}^2/M^2+u_0m_{\pi_1}^2/M^2}+\cdots\nonumber\\
&&=\frac{f_\pi m_\pi^2}{2\sqrt{2}}
\biggl\lbrace
\biggl[
\left(\frac{\partial\mathcal{A}_\perp}{\partial\alpha_3}-\frac{\partial\mathcal{A}_\perp}{\partial\alpha_2}\right)^{[\alpha_1]}
+\left(\frac{\partial\mathcal{V}_\parallel}{\partial\alpha_3}-\frac{\partial\mathcal{V}_\parallel}{\partial\alpha_2}\right)^{[\alpha_1]}
-\left(\frac{\partial\mathcal{V}_\perp}{\partial\alpha_3}-\frac{\partial\mathcal{V}_\perp}{\partial\alpha_2}\right)^{[\alpha_1]}
-\left(\frac{\partial\mathcal{A}_\perp}{\partial\alpha_3}-\frac{\partial\mathcal{A}_\perp}{\partial\alpha_1}\right)^{[\alpha_2]}
\nonumber\\
&&\phantom{=}+\left(\frac{\partial\mathcal{V}_\parallel}{\partial\alpha_3}-\frac{\partial\mathcal{V}_\parallel}{\partial\alpha_1}\right)^{[\alpha_2]}
-\left(\frac{\partial\mathcal{V}_\perp}{\partial\alpha_3}-\frac{\partial\mathcal{V}_\perp}{\partial\alpha_1}\right)^{[\alpha_2]}
+\mathcal{A}_\perp(\bar{u}_0,u_0,0)-\mathcal{A}_\perp(u_0,\bar{u}_0,0)+\mathcal{V}_\parallel(\bar{u}_0,u_0,0)
+\mathcal{V}_\parallel(u_0,\bar{u}_0,0)
\nonumber\\
&&\phantom{=}-\mathcal{V}_\perp(\bar{u}_0,u_0,0)-\mathcal{V}_\perp(u_0,\bar{u}_0,0)\biggl]M^4
+[\phi_\pi'(\bar{u}_0)-\phi_\pi'(u_0)]\frac{\langle g_s^2G^2\rangle}{m_\pi^2}M^2
+\int_{\bar{u}_0}^{u_0}\mathbb{B}(u)\,du\,\langle g_s^2G^2\rangle\biggl\rbrace\,,
\end{eqnarray}
\begin{eqnarray}\label{f1presumrule2}
&&f_{f_1}\tilde{f}_{\pi_1}m_{f_1}g_{f_1}^2 e^{\bar{u}_0m_{f_1}^2/M^2+u_0m_{\pi_1}^2/M^2}+\cdots\nonumber\\
&&=\sqrt{2}f_\pi m_\pi^2
\biggl\lbrace
\biggl[
\mathcal{A}_\parallel^{[\alpha_2,\bar{\alpha}_2]}-\mathcal{A}_\parallel^{[\alpha_1,\bar{\alpha}_1]}
+\mathcal{A}_\perp^{[\alpha_2,\bar{\alpha}_2]}-\mathcal{A}_\perp^{[\alpha_1,\bar{\alpha}_1]}
-\mathcal{V}_\parallel^{[\alpha_2,\bar{\alpha}_2]}-\mathcal{V}_\parallel^{[\alpha_1,\bar{\alpha}_1]}
-\mathcal{V}_\perp^{[\alpha_2,\bar{\alpha}_2]}-\mathcal{V}_\perp^{[\alpha_1,\bar{\alpha}_1]}
\nonumber\\
&&\phantom{=}+\mathcal{A}_\parallel^{[\alpha_1,u_0]}-\mathcal{A}_\parallel^{[\alpha_2,u_0]}
+\mathcal{A}_\perp^{[\alpha_1,u_0]}-\mathcal{A}_\perp^{[\alpha_2,u_0]}
+\mathcal{V}_\parallel^{[\alpha_1,u_0]}+\mathcal{V}_\parallel^{[\alpha_2,u_0]}
+\mathcal{V}_\perp^{[\alpha_1,u_0]}+\mathcal{V}_\perp^{[\alpha_2,u_0]}
+{1\over2}(\mathcal{V}_\perp^{[\alpha_1]}+\mathcal{V}_\perp^{[\alpha_2]})
\nonumber\\
&&\phantom{=}+\frac{1-4u_0}{2}(\mathcal{A}_\perp^{[\alpha_2]}-\mathcal{A}_\perp^{[\alpha_1]})
-\frac{1-2u_0}{2}(\mathcal{V}_\parallel^{[\alpha_1]}+\mathcal{V}_\parallel^{[\alpha_2]})\biggl]M^2
+\frac{\sqrt{2}}{2}u_0[\phi_\pi(u_0)+\phi_\pi(\bar{u}_0)]\frac{\langle g_s^2G^2\rangle}{m_\pi^2}
\nonumber\\
&&\phantom{=}+\frac{u_0\bar{u}_0}{72}\int_{\bar{u}_0}^{u_0}\mathbb{B}(u)\,du\frac{\langle g_s^2G^2\rangle}{M^2}
\biggl\rbrace\,.
\end{eqnarray}

\begin{figure}[!htb]
\begin{minipage}[t]{0.5\linewidth}
\centering
\captionstyle{flushleft}
\includegraphics[width=3in]{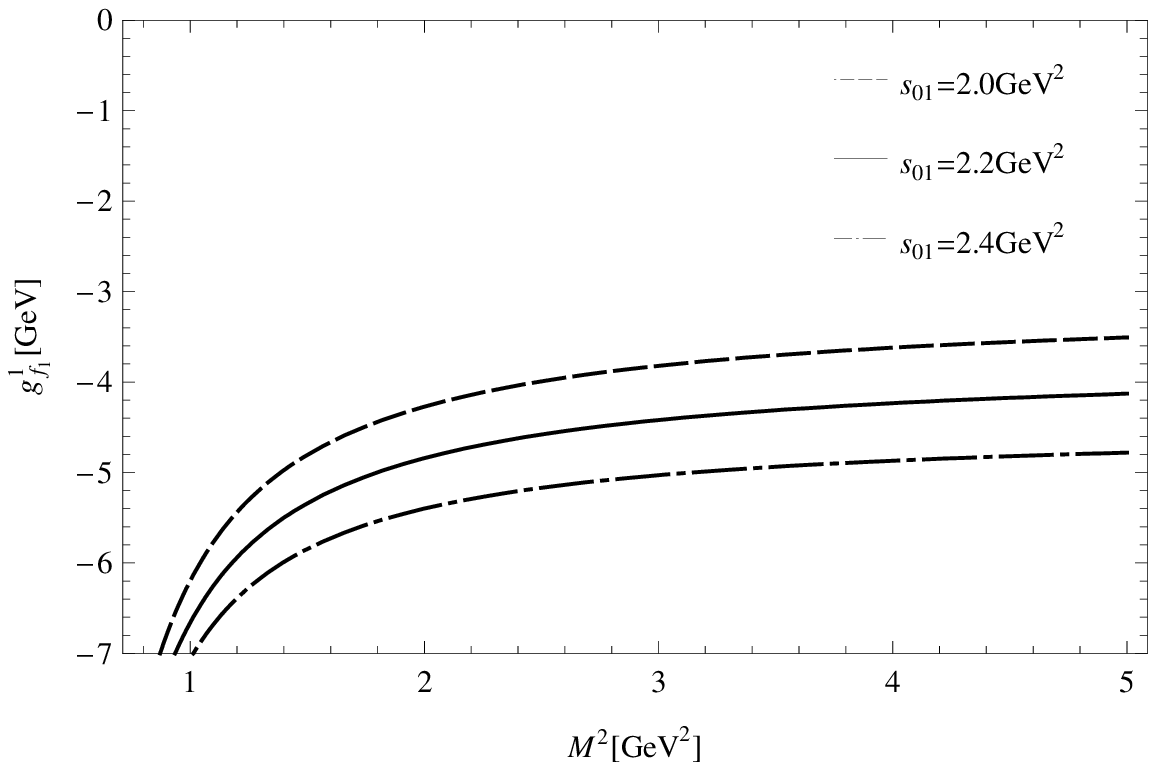}
\setcaptionwidth{3in}
\caption{The sum rule for $g_{f_1}^1$ with $m_{\pi_1}=1.6\ \text{GeV}$, $1.7<M^2<2.1\ \text{GeV}^2$,
and $s_{01}=2.0,2.2,2.4\ \text{GeV}^2$.} \label{fig:CCF1PiSm16}
\end{minipage}%
\begin{minipage}[t]{0.5\linewidth}
\centering
\captionstyle{flushleft}
\includegraphics[width=3in]{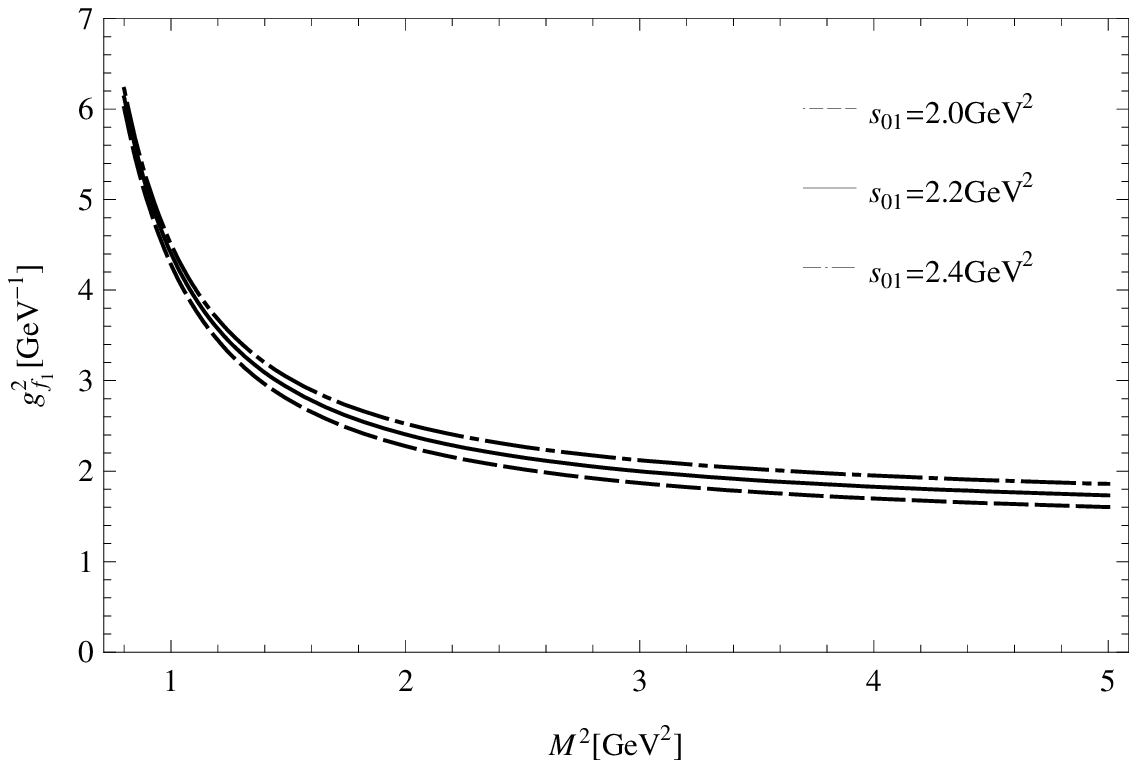}
\setcaptionwidth{3in}
\caption{The sum rule for $g_{f_1}^2$ with $m_{\pi_1}=1.6\ \text{GeV}$, $2.3<M^2<2.7\ \text{GeV}^2$,
and $s_{01}=2.0,2.2,2.4\ \text{GeV}^2$.} \label{fig:CCF1PiDm16}
\end{minipage}
\end{figure}

Here we omitted the terms $\sim\mathcal{O}(m_\pi^4)$.
After subtracting the continuum contribution, we get the sum rules for $g_{f_1}^1$ and $g_{f_1}^2$,
which are plotted with $s_{01}=2.0,2.2,2.4\ \text{GeV}^2$ in Figs. \ref{fig:CCF1PiSm16}-\ref{fig:CCF1PiDm16}.

The derivation of the sum rules for $g_{b_1}^1$ and $g_{b_1}^2$ is almost the same as mentioned above.
The interpolating current for the $b_1$ meson reads
\begin{eqnarray}
J^{b_1}_\mu(x)=\bar{d}(x)\overleftrightarrow{\partial}_\mu\gamma_5u(x)\,,
\end{eqnarray}
where $\overleftrightarrow{\partial}_\mu\equiv\overrightarrow{\partial}_\mu-\overleftarrow{\partial}_\mu$,
and we define $\langle 0|J^{b_1}_\mu(0)|b_1(k,\lambda)\rangle=f_{b_1}\epsilon^\lambda_\mu$.
There is another possible interpolating current for $b_1$:
$J'_\mu(x)=\bar{d}(x)\sigma_{\mu\nu}u(x)$,
which couples to $b_1$ through
$\langle 0|J'_{\mu\nu}(0)|b_1(k,\lambda)\rangle
=if^T_{b_1}\varepsilon_{\mu\nu\rho\sigma}\epsilon^\rho_\lambda k^\sigma$.
The same current couples to the $\rho$ meson through
$\langle 0|J'_{\mu\nu}(0)|\rho(k,\lambda)\rangle
=if^T_\rho(\epsilon_\mu^\lambda k_\nu-\epsilon_\nu^\lambda k_\mu)$.
Unfortunately, the Lorentz structure of these two coupling will mix with each other in the correlation function
\begin{eqnarray}
\Pi'_{\mu\nu\beta}(k^2,p^2)
=i\int d^4 xe^{ik\cdot x}\langle \pi(q)|T\{J'_{\mu\nu}(x)J^{\pi_1\dagger}_\beta(0)\}|0\rangle\,,
\end{eqnarray}
so that we are unable to separate the contribution of the $b_1$
part from that of the $\rho$ part. This is the consideration
behind our choice of the interpolating current $J^{b_1}_{\mu}$ for
the $b_1$ meson instead of the tensor one.

\begin{figure}[!htb]
\begin{minipage}[t]{0.5\linewidth}
\centering
\captionstyle{flushleft}
\includegraphics[width=3in]{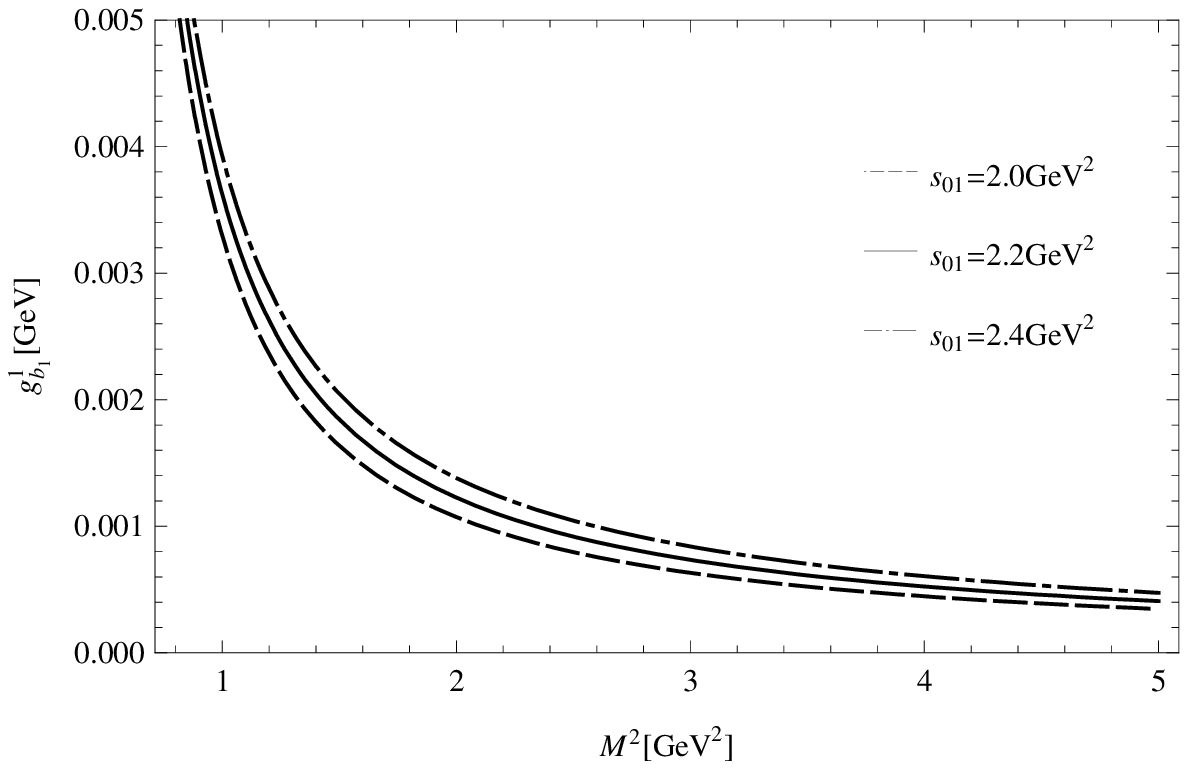}
\setcaptionwidth{3in}
\caption{The sum rule for $g_{b_1}^1$ with $m_{\pi_1}=1.6\ \text{GeV}$
and $s_{01}=2.0,2.2,2.4\ \text{GeV}^2$.
There is no stable working interval of $M^2$ for the sum rule.} \label{fig:CCB1PiSm16}
\end{minipage}%
\begin{minipage}[t]{0.5\linewidth}
\centering
\captionstyle{flushleft}
\includegraphics[width=3in]{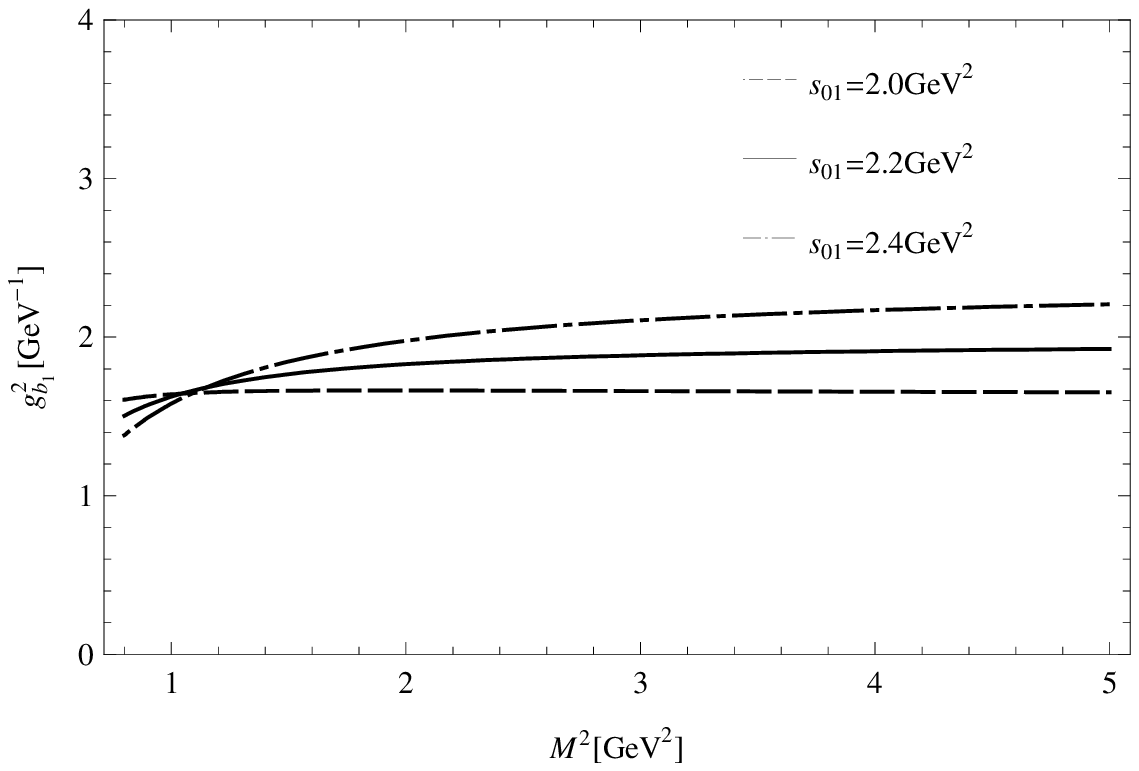}
\setcaptionwidth{3in}
\caption{The sum rule for $g_{b_1}^2$ with $m_{\pi_1}=1.6\ \text{GeV}$
and $s_{01}=2.0,2.2,2.4\ \text{GeV}^2$.
There is no stable working interval of $M^2$ for the sum rule.} \label{fig:CCB1PiDm16}
\end{minipage}
\end{figure}

The sum rules for $g_{b_1}^1$ and $g_{b_1}^2$ read as
\begin{eqnarray}\label{b1presumrule1}
f_{b_1}\tilde{f}_{\pi_1}g_{b_1}^1 e^{\bar{u}_0m_{b_1}^2/M^2+u_0m_{\pi_1}^2/M^2}+\cdots
=\frac{f_\pi m_\pi^2}{216\sqrt{2}(m_u+m_d)}\left[\phi_\sigma'(u_0)-\phi_\sigma'(\bar{u}_0)\right]M^2\langle g_s^2G^2\rangle\,,
\end{eqnarray}
\begin{eqnarray}\label{b1presumrule2}
&&f_{b_1}\tilde{f}_{\pi_1}g_{b_1}^2 e^{\bar{u}_0m_{b_1}^2/M^2+u_0m_{\pi_1}^2/M^2}+\cdots\nonumber\\
&&=\frac{f_\pi m_\pi^2}{108\sqrt{2}(m_u+m_d)}
\biggl\lbrace
216u_0\biggl[\mathcal{T}(u_0,\bar{u}_0,0)+\mathcal{T}(\bar{u}_0,u_0,0)
+\left(\frac{\partial\mathcal{T}}{\partial\alpha_3}-\frac{\partial\mathcal{T}}{\partial\alpha_2}\right)^{[\alpha_1]}
+\left(\frac{\partial\mathcal{T}}{\partial\alpha_3}-\frac{\partial\mathcal{T}}{\partial\alpha_1}\right)^{[\alpha_2]}\nonumber\\
&&\phantom{=}-\mathcal{T}^{[\alpha_1]}-\mathcal{T}^{[\alpha_2]}\biggl]M^4
+\biggl[(1-2u_0)[\phi_\sigma(u_0)+\phi_\sigma(\bar{u}_0)]+u_0\bar{u}_0[\phi_\sigma'(u_0)-\phi_\sigma'(\bar{u}_0)]\biggl]\langle g_s^2G^2\rangle
\biggl\rbrace\,.
\end{eqnarray}
They are plotted in Fig. \ref{fig:CCB1PiSm16}-\ref{fig:CCB1PiDm16} after the continuum substraction.

The adopted values of the parameters $f_{f_1}=0.17\ \text{GeV}$ and $f_{b_1}=0.18\ \text{GeV}^3$
are obtained from Eq. (4.52) and Eq. (A.20) in Ref. \cite{b1parameter}, respectively.
The extracted values of $g_{f_1}^1$, $g_{f_1}^2$, $g_{b_1}^1$, and $g_{b_1}^2$ are collected in Table \ref{tablef1cc}.

\begin{table}[!htb]
\begin{center}
\setlength\extrarowheight{6pt}
\begin{tabular}{ccccccc}
\hline
$m_{\pi_1}[\text{GeV}]$                & $1.6$                      & $1.8$                      & $2.0$                      & $M^2[\text{GeV}^2]$   \\
$g_{f_1}^1[\text{GeV}]$              & $\quad-4.2\sim-5.6\quad$   & $\quad-4.4\sim-5.9\quad$   & $\quad-4.6\sim-6.1\quad$   & $\quad1.7\sim2.1\quad$ \\
$g_{f_1}^2[\text{GeV}^{-1}]$         & $\quad2.0\sim2.4\quad$     & $\quad2.1\sim2.5\quad$     & $\quad2.3\sim2.7\quad$     & $\quad2.3\sim2.7\quad$ \\
$g_{b_1}^1[\text{GeV}]$              & $\ll 0.1$                  & $\ll 0.1$                  & $\ll 0.1$                  & - \\
$g_{b_1}^2[\text{GeV}^{-1}]$         & $\quad1.9\quad$            & $\quad1.8\quad$            & $\quad1.6\quad$            & - \\
\hline
\end{tabular}
\end{center}
\caption{The numerical values of $g_{f_1}^1$, $g_{f_1}^2$, $g_{b_1}^1$, and $g_{b_1}^2$.
The working intervals of the Borel parameter $M^2$ are listed in the right column.
Here ``-'' indicates the nonexistence of a stable working interval for the corresponding sum rule.
The values presented in these cases are determined by $M^2=2.0\ \text{GeV}^2$.
The range of the threshold is $2.0<s_{01}<2.4\ \text{GeV}^2$.
}
\label{tablef1cc}
\end{table}

\section{Sum rules for the $\pi_1\rightarrow \eta\pi,\eta'\pi$ coupling constants}\label{gammacoupling}
The interpolating currents for $\eta$ and $\eta'$ read as
\begin{eqnarray}
J^{\eta}&=&J^{\eta_8}\cos\theta-J^{\eta_1}\sin\theta\,,\nonumber\\
J^{\eta'}&=&J^{\eta_8}\sin\theta +J^{\eta_1}\cos\theta\,,
\end{eqnarray}
where
\begin{eqnarray}
J^{\eta_8}(x)&=&\frac{1}{\sqrt{6}}\left[\bar{u}(x)i\gamma_5 u(x)
+\bar{d}(x)i\gamma_5 d(x)-2\bar{s}(x)i\gamma_5 s(x)\right]\,,\nonumber\\
J^{\eta_1}(x)&=&\frac{1}{\sqrt{3}}\left[\bar{u}(x)i\gamma_5 u(x)
+\bar{d}(x)i\gamma_5 d(x)+\bar{s}(x)i\gamma_5 s(x)\right]\,,
\end{eqnarray}
and $\theta=-19^\circ$ is the mixing angle between $\eta_8$ and the SU(3) singlet $\eta_1$.
Their coupling to $\eta$ and $\eta'$ are
$\langle 0|J^{\eta}(0)|\eta(k)\rangle=\lambda_\eta$
and $\langle 0|J^{\eta'}(0)|\eta'(k)\rangle=\lambda_{\eta'}$, respectively.
The correlation function involved in our calculation is
\begin{eqnarray}
\Pi_{\eta/\eta'}(k^2,p^2)=i\int d^4 xe^{ik\cdot x}\langle \pi(q)|T\{J^{\eta/\eta'}(x)J^{\pi_1\dagger}_\beta(0)\}|0\rangle
&=&q^\beta G_{\eta/\eta'}(k^2,p^2)+\cdots\,,
\end{eqnarray}
The sum rules derived for $g_\eta$ and $g_{\eta'}$ are
\begin{eqnarray}\label{etapresumrule}
&&\lambda_\eta \tilde{f}_{\pi_1}g_\eta e^{\bar{u}_0m_\eta^2/M^2+u_0m_{\pi_1}^2/M^2}+\cdots\nonumber\\
&&=\frac{\cos\theta f_\pi m_\pi^2}{216\sqrt{2}(m_u+m_d)}
\biggl\lbrace
216\biggl[\left(\frac{\partial\mathcal{T}}{\partial\alpha_3}-\frac{\partial\mathcal{T}}{\partial\alpha_2}\right)^{[\alpha_1]}
+\left(\frac{\partial\mathcal{T}}{\partial\alpha_3}-\frac{\partial\mathcal{T}}{\partial\alpha_1}\right)^{[\alpha_2]}
+\mathcal{T}(u_0,\bar{u}_0,0)+\mathcal{T}(\bar{u}_0,u_0,0)\biggl]M^4
\nonumber\\
&&\phantom{=}+\bar{u}_0[\phi_\sigma'(u_0)-\phi_\sigma'(\bar{u}_0)]\langle g_s^2G^2\rangle
\biggl\rbrace\,,
\end{eqnarray}
\begin{eqnarray}\label{etaprimepresumrule}
&&\lambda_{\eta'} \tilde{f}_{\pi_1}g_\eta e^{\bar{u}_0m_{\eta'}^2/M^2+u_0m_{\pi_1}^2/M^2}+\cdots\nonumber\\
&&=\frac{\sin\theta f_\pi m_\pi^2}{216\sqrt{2}(m_u+m_d)}
\biggl\lbrace
216\biggl[\left(\frac{\partial\mathcal{T}}{\partial\alpha_3}-\frac{\partial\mathcal{T}}{\partial\alpha_2}\right)^{[\alpha_1]}
+\left(\frac{\partial\mathcal{T}}{\partial\alpha_3}-\frac{\partial\mathcal{T}}{\partial\alpha_1}\right)^{[\alpha_2]}
+\mathcal{T}(u_0,\bar{u}_0,0)+\mathcal{T}(\bar{u}_0,u_0,0)\biggl]M^4
\nonumber\\
&&\phantom{=}+\bar{u}_0[\phi_\sigma'(u_0)-\phi_\sigma'(\bar{u}_0)]\langle g_s^2G^2\rangle
\biggl\rbrace\,.
\end{eqnarray}

We take $m_\eta=0.547\ \text{GeV}$, $\lambda_\eta=0.23\ \text{GeV}^2$,
$m_{\eta^\prime}=0.958\ \text{GeV}$, and $\lambda_{\eta^\prime}=0.33\ \text{GeV}^2$
in our numerical analysis \cite{etaparameter}.
The extracted values of $g_\eta$ and $g_{\eta'}$ are presented here:

\begin{center}
\setlength\extrarowheight{8pt}
\begin{tabular}{ccccc}
\hline
$m_{\pi_1}[\text{GeV}]$  & $1.6$               & $1.8$               & $2.0$               & $M^2[\text{GeV}^2]$   \\
$g_\eta$               & $\quad0.45\quad$    & $\quad0.45\quad$    & $\quad0.42\quad$    & -   \\
$g_{\eta'}$            & $\quad-0.16\quad$   & $\quad-0.15\quad$   & $\quad-0.15\quad$   & -   \\
\hline
\end{tabular}
\end{center}

\section{Partial Decay Widths}\label{decaywidth}

It is straightforward to calculate the partial widths of $\pi_1$ with the extracted coupling constants.
The formulae for these partial widths are given in Eq. (\ref{widthformulae}).
\begin{eqnarray}\label{widthformulae}
&&\Gamma_{\pi_1^0\rightarrow \rho^+\pi^-}
=\frac{g_\rho^2}{96\pi m_{\pi_1}^3}|\vec{q}_\pi|^3\,,\nonumber\\
&&\Gamma_{\pi_1^0\rightarrow f_1\pi^0}
=\frac{1}{24\pi m_{\pi_1}^2}\Big[(g^1_{f_1})^2(3+\frac{|\vec{q}_\pi|^2}{m^2_{f_1}})|\vec{q}_\pi|
+(g^2_{f_1})^2\frac{m^2_{\pi_1}}{m^2_{f_1}}|\vec{q}_\pi|^5
+2g^1_{f_1}g^2_{f_1}\frac{m_{\pi_1}}{m^2_{f_1}}\sqrt{m^2_{f_1}+|\vec{q}_\pi|^2}|\vec{q}_\pi|^3\Big]\,,\nonumber\\
&&\Gamma_{\pi_1^0\rightarrow b_1^+\pi^-}
=\Gamma_{\pi_1^0\rightarrow f_1\pi^0}(g_{f_1}^1\rightarrow g_{b_1}^1, g_{f_1}^2\rightarrow g_{b_1}^2, m_{f_1}\rightarrow m_{b_1})\,,\nonumber\\
&&\Gamma_{\pi_1^0\rightarrow \eta\pi^0}
=\frac{g_\eta^2}{192\pi m_{\pi_1}^3}|\vec{q}_\pi|^3\,,\nonumber\\
&&\Gamma_{\pi_1^0\rightarrow \eta'\pi^0}
=\Gamma_{\pi_1^0\rightarrow \eta\pi^0}(g_\eta\rightarrow g_{\eta'}, m_\eta\rightarrow m_{\eta'})\,.
\end{eqnarray}
For completeness, we make rough estimates of the decay widths of
some decay modes with the given coupling constants, although the
corresponding coupling constants are extracted from sum rules
without a stable working interval. We collected our results in
Table \ref{tablewidthspi}, together with the results obtained
using other phenomenological models, e.g. Ref. \cite{PSS} (PSS)
and Ref. \cite{IKP} (IKP). Here we also reproduce Table XIII of
Ref. \cite{meyer} to present the existing experimental results on
the total decay width of $\pi_1$ in Table \ref{tablewidthspiexp} in
order to compare with our predictions.

\begin{table}[!htb]
\begin{center}
\setlength\extrarowheight{3pt}
\begin{tabular}{c|lllllllllllllll}
\hline
$m_{\pi_1}[\text{GeV}]$ &  \multicolumn{3}{c}{$1.6$}     &   \multicolumn{3}{c}{$1.8$}         &  \multicolumn{4}{c}{$2.0$}                        \\
                      & \ IKP  &   PSS  &  this work   & $\quad$IKP  & PSS    &  this work   & $\quad$IKP &  PSS   & Lattice     &  this work     \\
                      & \ \cite{IKP}&\cite{PSS}&       & $\quad$\cite{IKP}&\cite{PSS}&       & $\quad$\cite{IKP}&\cite{PSS}&\cite{lattwidth}&     \\
$\rho\pi$             & \ 8    & 9      & $73\sim120$  & $\quad$12   & 13     & $138\sim222$ & $\quad$    & 16     &             & $216\sim370$   \\
$f_1\pi$              & \ 14   & 5      & $69\sim122$  & $\quad$21   & 9      & $96\sim175$  & $\quad$    & 10.2   & $90\pm60$   & $109\sim195$   \\
$b_1\pi$              & \ 59   & 24     & $0.14$       & $\quad$62   & 38     & $1.2$        & $\quad$    & 43     & $400\pm120$ & $3.7$          \\
$\eta\pi$             & \ 0    & 0      & $0.36$       & $\quad$.02  & .02    & $0.44$       & $\quad$    & .02    &             & $0.45$         \\
$\eta'\pi$            & \ 0    & 0      & $0.02$       & $\quad$0    & .01    & $0.02$       & $\quad$    & .01    &             & $0.03$         \\
\hline
\end{tabular}
\end{center}
\caption{The partial widths of the single-pion channels of $\pi_1$
in units of MeV, where IKP and PSS refer to the methods used in
Ref. \cite{IKP} and Ref. \cite{PSS}, respectively.
The partial widths for the channels $b_1\pi$ and $f_1\pi$ cited from IKP and PSS are simply the sum of their $S$-wave and $D$-wave widths.}
\label{tablewidthspi}         
\end{table}

\begin{table}[!htb]\centering
\setlength\extrarowheight{5pt}
\begin{tabular}{cllcc} \hline\hline
Mode & Mass (GeV) & Width (GeV) & Experiment & Reference \\
\hline
$\rho\pi$              & $1.593\pm 0.08$  & $0.168\pm 0.020$     & E852    & \cite{1600E8521} \\
$\eta^{\prime}\pi$     & $1.597\pm 0.010$ & $0.340\pm 0.040$     & E852    & \cite{1600E8522} \\
$f_{1}\pi$             & $1.709\pm 0.024$ & $0.403\pm 0.080$     & E852    & \cite{1600E8523} \\
$b_{1}\pi$             & $1.664\pm 0.008$ & $0.185\pm 0.025$     & E852    & \cite{1600E8524} \\
$b_{1}\pi$             & $1.58\pm 0.03$   & $0.30\pm 0.03$       & VES     & \cite{VES1} \\
$b_{1}\pi$             & $1.61\pm 0.02$   & $0.290\pm 0.03$      & VES     & \cite{1600VES2} \\
$b_{1}\pi$             & $\sim 1.6$       & $\sim 0.33$          & VES     & \cite{VES2} \\
$b_{1}\pi$             & $1.56\pm 0.06$   & $0.34\pm 0.06$       & VES     & \cite{VES3} \\
$f_{1}\pi$             & $1.64\pm 0.03$   & $0.24\pm 0.06$       & VES     & \cite{VES3} \\
$\eta^{\prime}\pi$     & $1.58\pm 0.03$   & $0.30\pm 0.03$       & VES     & \cite{VES1} \\
$\eta^{\prime}\pi$     & $1.61\pm 0.02$   & $0.290\pm 0.03$      & VES     & \cite{VES2} \\
$\eta^{\prime}\pi$     & $1.56\pm 0.06$   & $0.34\pm 0.06$       & VES     & \cite{VES3} \\
$b_{1}\pi$             & $\sim 1.6$       & $\sim 0.23$          & CBAR    & \cite{1600Crystal} \\
$\rho\pi$              & $1.660\pm 0.010$ & $0.269\pm 0.021$     & COMPASS & \cite{compass} \\
all                    & $1.662^{+0.015}_{-0.011}$ & $0.234\pm 0.050$ & PDG & \cite{PDG} \\
\hline\hline
\end{tabular}
\caption[]{\label{tablewidthspiexp}Reported masses and widths of the $\pi_{1}(1600)$ from the
E852 experiment, the VES experiment and the COMPASS experiment. The PDG average from
2008 is also reported. (Table reproduced from Ref. \cite{meyer}.) }
\end{table}

The numerical values of the coupling constants of the modes
$\pi_1\rightarrow \rho\pi, f_1\pi$ are stable with the variation of
$m_{\pi_1}$. As a consequence, the partial widths of these two modes
increase rapidly with $m_{\pi_1}$ due to their enlarged two-body
phase spaces.

Apparently, our results on the partial width of $\pi_1$ are quite
different from those obtained using Lattice QCD \cite{lattwidth}
and other phenomenological approaches such as the IKP and PSS flux
tube model. In the flux tube model, the $\pi_1$ coupling to the
two $S$-wave mesons is suppressed, leading to a small branch ratio
of the mode $\pi_1\rightarrow\rho\pi$. One flux tube model
prediction \cite{fluxtubewidth} for widths for $\pi_1$ with
$m_{\pi_1}=2.0\ \text{GeV}$ is (in MeV)
\begin{eqnarray}
\pi f_1 : \pi b_1 : \pi \rho : \eta \pi : \eta' \pi \;
= \; 60 : 170 : 5\sim20 : 0\sim10 : 0\sim10\;.
\end{eqnarray}
The quenched lattice QCD simulation also predicted a large width
of the channel $b_1\pi$ \cite{lattwidth}, although the linear
extrapolation approximation adopted there may lead to an
overestimated width for this channel, as pointed out in Ref.
\cite{fluxtubelattice}. However, the $b_1\pi$ channel is severely
suppressed in our calculation since the numerical value of
$g_{b_1}^1$ is found to be extremely small. As far as the channels
$\eta\pi$ and $\eta'\pi$ are concerned, Chung, Klempt, and Korner
argued that the channel $\pi_1\rightarrow\eta\pi$ is forbidden due
to the requirement of Bose symmetry and $J^{PC}$ conservation in
the limit that the $\eta$ is a pure SU(3) octet
\cite{etapiforbidden}. The tiny mixing between $\eta_8$ and
$\eta_1$ should not reverse the widths of these two channels. In
contrast, the width of $\eta\pi$ is at least one order of
magnitude larger than that of the channel $\eta'\pi$ in our
calculation. Experimentally, the relative branching ratios for
$\pi_1(1600)$'s three channels $b_1\pi$, $\eta'\pi$, and $\rho\pi$
are \cite{1600VES2}
\begin{eqnarray}
b_1\pi : \eta'\pi : \rho\pi \; = \; 1 : 1 \pm 0.3 : 1.5 \pm 0.5.
\end{eqnarray}
In a summary on the VES results,
Amelin \cite{VES3} obtained the relative branch ratios for the $\pi_1(1600)$ as follows:
\begin{eqnarray}
b_1\pi : f_1\pi : \rho\pi : \eta'\pi \; = \; 1.0\pm .3 : 1.1 \pm .3 : < .3 : 1.
\end{eqnarray}

\section{Strong decays of the isoscalar $1^{-+}$ hybrid state}\label{isoscalar}

Now we consider the strong decays of $\tilde{\pi}_1$, the isoscalar parter of $\pi_1$.
We notice that $I^GJ^{PC}$ conservation restricts the possible decay channels to
the $S$-wave $\tilde{\pi}_1\rightarrow a_1(1260)\pi$, $f_1(1285)\eta$,
and the $P$-wave $\tilde{\pi}_1\rightarrow \eta\eta'$, $\pi(1300)\pi$, $\eta(1295)\eta$.
The partial widths of the three $P$-wave channels
are supposed to be relatively small due to their small phase spaces.
In addition, the channel $\tilde{\pi}_1\rightarrow f_1\eta$
is kinematically forbidden if the mass of $\tilde{\pi}_1$ is smaller than $1.83\ \text{GeV}$.
Hence the dominant decay mode of $\tilde{\pi}_1$ is $\tilde{\pi}_1\rightarrow a_1\pi$.

The interpolating currents for the $\tilde{\pi}_1$ meson and the $a_1$ meson are
\begin{eqnarray}
J^{\tilde{\pi}_1}_\mu(x)&=&\frac{1}{\sqrt{2}}\left[\bar{u}(x)\frac{\lambda^a}{2}g_sG^a_{\mu\nu}(x)\gamma^\nu
u(x)
+\bar{d}(x)\frac{\lambda^a}{2}g_sG^a_{\mu\nu}(x)\gamma^\nu d(x)\right]\,,\nonumber\\
J^{a_1}_\mu(x)&=&\frac{1}{\sqrt{2}}\left[\bar{u}(x)\gamma_\mu\gamma_5
u(x) -\bar{d}(x)\gamma_\mu\gamma_5 d(x)\right]\,.
\end{eqnarray}
We define the coupling constants $g_{a_1}^1$ and $g_{a_1}^2$,
similar to the coupling constants $g_{f_1}^1$ and $g_{f_1}^2$ of
the channel $\pi_1\rightarrow f_1\pi$. Apparently, the sum rules for
$g_{a_1}^1$ and $g_{a_1}^2$ before the continuum subtraction are
similar to Eq. (\ref{f1presumrule1}) and (\ref{f1presumrule2}),
respectively. The only difference between them lies in the
hadron-level parameters, namely the masses and the overlapping
amplitudes of the mesons involved in these two decay modes. Hence
it is plausible to write down $g_{a_1}^1\approx g_{f_1}^1$ and
$g_{a_1}^2\approx g_{f_1}^2$ if we simply adopt
$\tilde{f}_{\tilde{\pi}_1}=\tilde{f}_{\pi_1}=0.15\ \text{GeV}^4$ and
$f_{a_1}=f_{f_1}=0.17\ \text{GeV}$ in our numerical analysis. This
leads to the estimate of the partial width of the mode
$\tilde{\pi}_1\rightarrow a_1\pi$: $\Gamma_{\tilde{\pi}_1\rightarrow a_1\pi}\approx
3\Gamma_{\pi_1\rightarrow f_1\pi}$. Here the factor $3$ comes from
the difference between the two channels' final states, namely
$\tilde{\pi}_1\rightarrow a_1^+\pi^-,a_1^-\pi^+,a_1^0\pi^0$ versus
$\pi_1^0\rightarrow f_1\pi^0$.

If we denote the right side of the sum rules for $g_{f_1}^1$
in Eq. (\ref{f1presumrule1}) and $g_{f_1}^2$ in Eq. (\ref{f1presumrule2})
as $R_{f_1}^1$ and $R_{f_1}^1$, respectively,
then we have the following sum rules for $g_{f_1\eta}^1$ and $g_{f_1\eta}^2$:
\begin{eqnarray}
f_{f_1}\tilde{f}_{\tilde{\pi}_1}m_{f_1}g_{f_1\eta}^1 e^{\bar{u}_0m_{f_1}^2/M^2+u_0m_{\tilde{\pi}_1}^2/M^2}+\cdots
&=&\frac{1}{\sqrt{3}}R_{f_1}^1(\pi\rightarrow\sigma, f_\pi\rightarrow f_\sigma, m_\pi\rightarrow m_\sigma,\cdots)\,, \nonumber\\
f_{f_1}\tilde{f}_{\tilde{\pi}_1}m_{f_1}g_{f_1\eta}^2 e^{\bar{u}_0m_{f_1}^2/M^2+u_0m_{\tilde{\pi}_1}^2/M^2}+\cdots
&=&\frac{1}{\sqrt{3}}R_{f_1}^2(\pi\rightarrow\sigma, f_\pi\rightarrow f_\sigma, m_\pi\rightarrow m_\sigma,\cdots)\,,
\end{eqnarray}
where we have ignored the SU(3) singlet octet mixing in the $\eta\eta'$ system
and the factor $e^{u_0\bar{u}_0m_\eta^2/M^2}$ because $m_\eta^2/M^2<0.1$ in our calculations.

\begin{figure}[!htb]
\begin{minipage}[t]{0.5\linewidth}
\centering
\captionstyle{flushleft}
\includegraphics[width=3in]{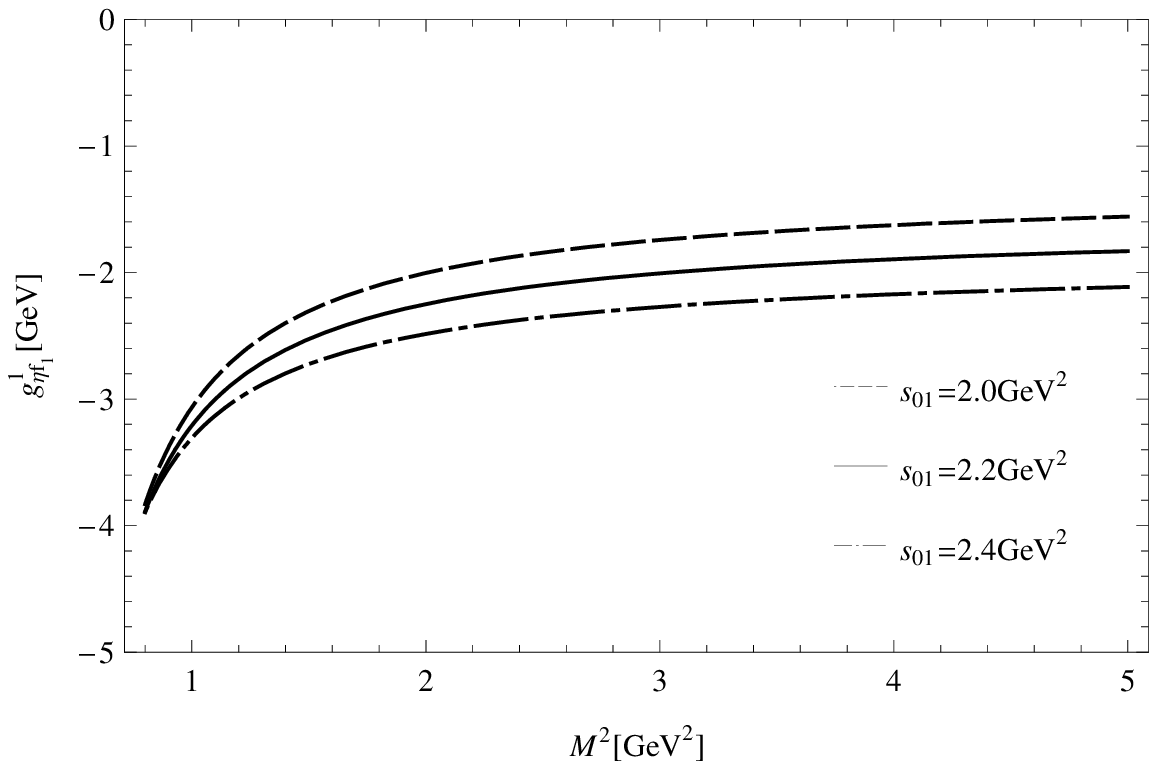}
\setcaptionwidth{3in}
\caption{The sum rule for $g_{f_1\eta}^1$ with $m_{\tilde{\pi}_1}=2.0\ \text{GeV}$, $2.6<M^2<3.0\ \text{GeV}^2$,
and $s_{01}=2.0,2.2,2.4\ \text{GeV}^2$.} \label{fig:CCF1EtaSm20}
\end{minipage}%
\begin{minipage}[t]{0.5\linewidth}
\centering
\captionstyle{flushleft}
\includegraphics[width=3in]{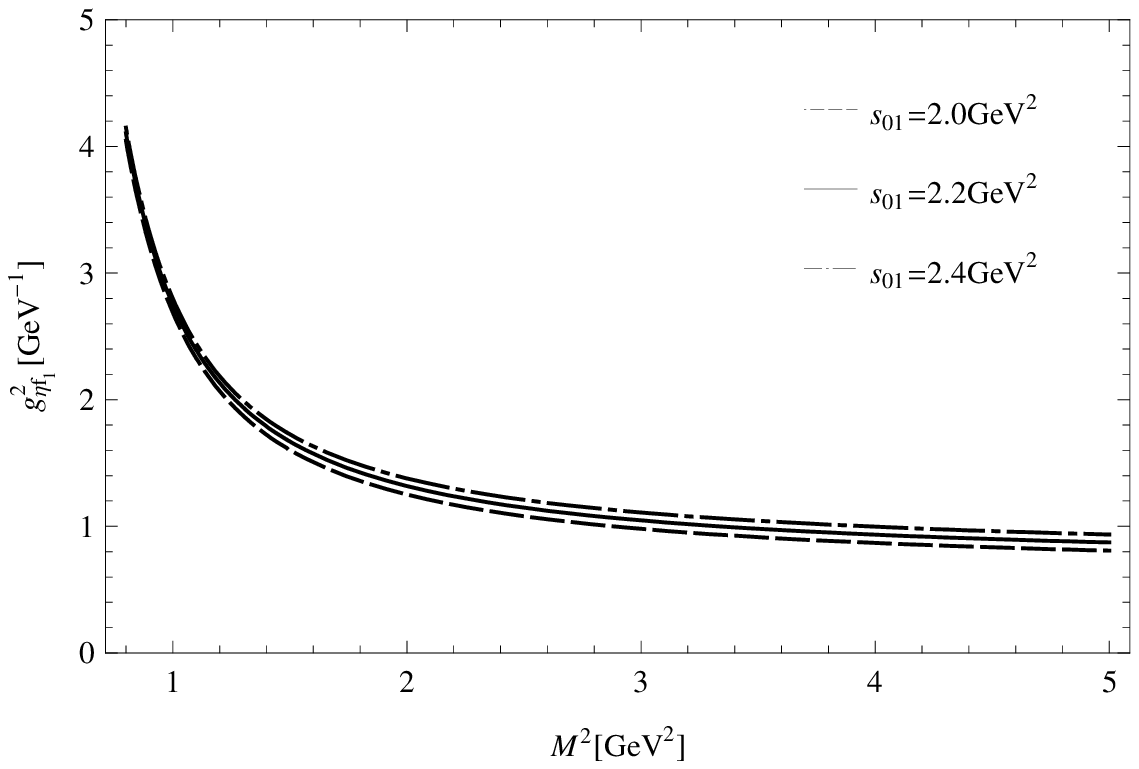}
\setcaptionwidth{3in}
\caption{The sum rule for $g_{f_1\eta}^2$ with $m_{\tilde{\pi}_1}=2.0\ \text{GeV}$, $2.6<M^2<3.0\ \text{GeV}^2$,
and $s_{01}=2.0,2.2,2.4\ \text{GeV}^2$.} \label{fig:CCF1EtaDm20}
\end{minipage}
\end{figure}

We adopt $f_\eta=130\ \text{MeV}$ and $\mu_\eta=1.47\ \text{GeV}$ \cite{pilcda}.
The mass of strange quark is taken to be $m_s=0.15\ \text{GeV}$.
The input parameters for the $\eta$ light-cone distribution amplitudes involved in our calculation
are as follows ($\mu=1\ \text{GeV}$) \cite{pilcda}:
\begin{center}
\setlength\extrarowheight{8pt}
\begin{tabular}{ccccccccccc}
\hline
  $a_2$\ \ &$\eta_3$\ \  &$\omega_3$\ \ &$\eta_4$\ \   &$\omega_4$\ \  &$h_{00}$\ \  &$v_{00}$\ \  &$a_{10}$\ \  &$v_{10}$\ \  &$h_{01}$\ \  &$h_{10}$\\
  $0.2$\ \ &$0.013$\ \   &$-3$\ \       &$0.5$\ \      &$0.2$\ \       &$-0.17$\ \   &$-0.17$\ \   &$0.17$\ \    &$0.26$\ \    &$0.15$\ \    &$0.38$  \\
\hline
\end{tabular}
\end{center}

The sum rules after the continuum substraction
are plotted in Fig. \ref{fig:CCF1EtaSm20}-\ref{fig:CCF1EtaDm20}.
Their numerical values for $2.0<s_{01}<2.4\ \text{GeV}^2$ are given here:

\begin{center}
\setlength\extrarowheight{8pt}
\begin{tabular}{ccccccc}
\hline
$m_{\tilde{\pi}_1}[\text{GeV}]$                    & $1.6$                    & $1.8$                    & $2.0$                    & $M^2[\text{GeV}^2]$   \\
$g_{f_1\eta}^1[\text{GeV}]$              & $\quad-1.6\sim-2.2\quad$ & $\quad-1.7\sim-2.3\quad$ & $\quad-1.7\sim-2.3\quad$ & $2.6\sim3.0$          \\
$g_{f_1\eta}^2[\text{GeV}^{-1}]$         & $\quad0.8\sim1.0\quad$   & $\quad0.9\sim1.1\quad$   & $\quad1.0\sim1.2\quad$   & $2.6\sim3.0$          \\
\hline
\end{tabular}
\end{center}

Our calculation of $\tilde{\pi}_1$'s widths are straightforward if we take advantage of the width formulae for $\pi_1\rightarrow f_1\pi$:
\begin{eqnarray}
&&\Gamma_{\tilde{\pi}_1^0\rightarrow a_1^+\pi^-}
=\Gamma_{\pi_1^0\rightarrow f_1\pi^0}(g_{f_1}^1\rightarrow g_{a_1}^1, g_{f_1}^2\rightarrow g_{a_1}^2, m_{f_1}\rightarrow m_{a_1}, m_{\pi_1}\rightarrow m_{\tilde{\pi}_1})\,,\nonumber\\
&&\Gamma_{\tilde{\pi}_1^0\rightarrow f_1\eta}
=\Gamma_{\pi_1^0\rightarrow f_1\pi^0}(g_{f_1}^1\rightarrow g_{f_1\eta}^1, g_{f_1}^2\rightarrow g_{f_1\eta}^2, m_{\pi_1}\rightarrow m_{\tilde{\pi}_1}, m_\pi\rightarrow m_\eta)\,.
\end{eqnarray}
The partial widths of $\tilde{\pi}_1$ are listed in Table \ref{scalarwidths}, along with the results obtained using other approaches.
As mentioned above, the only dominant channel for $\tilde{\pi}_1$ is $\tilde{\pi}_1\rightarrow a_1\pi$ whose width varies from $200\ \text{MeV}$ to $600\ \text{MeV}$,
depending heavily on the mass of $\tilde{\pi}_1$ we adopt.

\begin{table}[!htb]
\begin{center}
\setlength\extrarowheight{6pt}
\begin{tabular}{c|lllllllllllll}
\hline
$m_{\tilde{\pi}_1}[\text{GeV}]$ &  \multicolumn{3}{c}{$1.6$}       &  \multicolumn{3}{c}{$1.8$}          &   \multicolumn{3}{c}{$2.0$}           \\
                      &  \ \ IKP &   PSS  &  this work   & $\quad$IKP  & PSS    &  this work   &  $\quad$IKP &  PSS   &  this work     \\
                      & \ \cite{IKP}&\cite{PSS}&         & $\quad$\cite{IKP}&\cite{PSS}&       & $\quad$\cite{IKP}&\cite{PSS}&         \\
$a_1\pi$              &  \ \     &        & $207\sim366$ & $\quad72$   & $28.2$ & $288\sim525$ &  $\quad$    & $30.6$ & $327\sim585$   \\
$f_1\eta$             &  \ \ -   & -      & -            & $\quad$-    & -      &  -           &             & $8$    & $10\sim19$     \\
\hline
\end{tabular}
\end{center}
\caption{The partial widths of $\tilde{\pi}_1$ in units of MeV,
where ``-'' indicates that the corresponding channel is kinematically forbidden.}
\label{scalarwidths}         
\end{table}

\section{Search for the $1^{-+}$ state at BESIII}\label{bes}

Since the BESIII detector has an excellent photon resolution, it's
very interesting to search for the $1^{-+}$ state in the $J/\psi$
radiative process $J/\psi (\psi')\to h_{1,0} +\gamma$. The photon
spectrum peaks around $E_\gamma = {m_{J/\psi}^2-m_h^2\over 2
m_{J/\psi}}$ with a width ${m_h \Gamma_h \over m_{J/\psi}}\sim
(100-200)$ MeV. Such a process may be described by the following
effective Lagrangian
\begin{equation}
{\cal L}=c_0 F_{\mu\nu} \psi^\mu h_{1,0}^\nu  +c'_0 F_{\mu\nu}
\psi^{\nu\alpha} {h_{1,0}}^\mu_\alpha
\end{equation}
where $F_{\mu\nu}$ is the electromagnetic field strength tensor,
$\psi_\mu$ and $h_{1,0}^\mu$ are the $J/\psi$ and $1^{-+}$ field.
Naively one expects the above branching ratio to be around
$10^{-5}\sim 10^{-4}$.

The isovector $1^{-+}$ state $\pi_1$ can also be produced associated
with other hadrons X at BESIII through the process $J/\psi
(\psi')\to \pi_1 + X$. For the production of the neutral component
of $\pi_1$, the quantum numbers of X are $I^G=1^+, C=-$. Moreover,
$m_X\le m_{J/\psi}-m_{\pi_1}\sim 1.5 $ GeV. From the above
constraint, we get $X=\rho^0, b_1^0, \rho(1450)$ if it is a single
resonance or $X=\pi^+ \pi^-$ etc. Let's focus on the case
$X=\rho$. Such a production may be described by the following
effective Lagrangian
\begin{equation}
{\cal L}=c_1\psi_{\mu\nu} {\vec h}_1^\mu \cdot {\vec \rho}^\nu
+c_2 \psi_{\mu} {\vec h}_1^{\mu\nu} \cdot {\vec \rho}^\nu +c_3
\psi_{\mu} {\vec h}_1^{\nu} \cdot {\vec \rho}^{\mu\nu} + c_4
\psi_{\mu\nu} {\vec h}_1^{\nu\alpha} \cdot {\vec \rho}^\mu_\alpha
\end{equation}
Naively one expects the above branching ratio to be around
$10^{-4}\sim 10^{-3}$.

From Table \ref{tablewidthspi}, the dominant decay modes of the
isovector $1^{-+}$ meson are $\rho\pi, f_1\pi$. We urge our BESIII
colleagues to search for $\pi_1$ through the decay chain: $J/\psi
(\psi')\to \pi_1 +\gamma$, $\pi_1\to \rho\pi\to \pi^+\pi^-\pi^0$ or
$\pi_1\to f_1(1285)\pi^0$. $f_1(1285)$ is a narrow state with a
width of 24.3 MeV. The $f_1\pi^0$ mode is also useful in the
search of $\pi_1$, although $f_1(1285)$ mainly decays into multiple
particle final states $4\pi, \eta \pi\pi$. The other important
decay chain is $J/\psi (\psi')\to \pi_1 +\rho \to \rho+\rho+\pi\to
2(\pi^+\pi^-) \pi^0$. Once enough data are accumulated, one may
also try to look for $\pi_1$ in the $b_1\pi, \eta\pi, \eta'\pi$
modes.

The isoscalar $1^{-+}$ state $\tilde{\pi}_1$ can also be produced
associated with other hadrons X' at BESIII through the process
$J/\psi (\psi')\to \tilde{\pi}_1 + X'$. Now the quantum numbers of
X' are $I^G=0^-, C=-$. The possible candidates are $X'=\omega,
\phi, h_1(1170), \omega(1470), \pi^+\pi^-\pi^0$ etc. The
$\tilde{\pi}_1$ state mainly decays into $a_1\pi$. Search of
$\tilde{\pi}_1$ through the hadronic decay chain $J/\psi
(\psi')\to \tilde{\pi}_1 + \omega/\phi\to a_1+ \pi + \omega/\phi$
is challenging since it involves too many pions in the final
states. BESIII collaboration may also search for $\tilde{\pi}_1$
through the radiative decay chain: $J/\psi (\psi')\to
\tilde{\pi}_1 + \gamma\to a_1+ \pi + \gamma$.

\section{Conclusion}\label{summary}

We have studied the major strong decay modes of the $J^{PC}=1^{-+}$ hybrid mesons,
including the isovector and the isoscalar cases.
The coupling constants for these modes are extracted with the Light-cone QCD sum rule approach.
Most of the sum rules obtained are stable
with the variations of the Borel parameter $M^2$ and the continuum
threshold $s_{01}$. For the other sum rules, we can not find a
stable working interval of $M^2$.

Some possible sources of the errors in our calculation include the
inherent inaccuracy of LCQSR: the omission of the higher twist
terms in the OPE near the light-cone, the variation of the
coupling constant with the continuum threshold $s_{01}$ and the
Borel parameter $M^2$ in the working interval, the omission of the
higher conformal partial waves in the light-cone distribution
amplitudes of the pion or the $\eta$, and the uncertainty
in the parameters that appear in these light-cone distribution amplitudes.
The uncertainty in the overlapping amplitudes between
the interpolating currents and the corresponding final mesons is
another source of errors.
We merely give an estimated range for each involved coupling constant.
The uncertainty of $\tilde{f}_h$, which was not taken into consideration in this work,
may broaden these ranges significantly.
It's also understood that the $\alpha_s$
correction may turn out to be quite large.
We also omit the $\mathcal{O}(m_\pi^4)$ terms in the derivation of the sum rules for most of
the coupling constants involved in our calculation.

The partial widths of the $1^{-+}$ hybrid calculated from the
coupling constants extracted here are quite different from those
obtained using other methods like the flux tube model and Lattice
QCD etc. So far, the experimental data on the decay pattern of the
$1^{-+}$ hybrid is still not so accurate. We suggest the possible
search of the isovector and the isoscalar $1^{-+}$ hybrids in
$J/\psi (\psi')$ decay processes at BESIII.

\section*{Acknowledgments}

The authors thank Prof. K. T. Chao for useful discussions. This
project is supported by the National Natural Science Foundation of
China under Grants No. 10625521, No. 10721063, and the Ministry of
Science and Technology of China (2009CB825200).


\appendix

\section{The light-cone distribution amplitudes of the pion}\label{appendixLCDA}

The 2-particle distribution amplitudes of the $\pi$ meson are defined as \cite{pilcda}
\begin{eqnarray}
\langle 0 | \bar u(z) \gamma_\mu\gamma_5 d(-z) |
  \pi^-(P)\rangle
& = & i f_\pi p_\mu \int_0^1 du\, e^{i\xi pz} \, \phi_\pi(u) +
  \frac{i}{2}\, f_\pi m^2\, \frac{1}{pz}\, z_\mu \int_0^1 du \,
  e^{i\xi pz} g_\pi(u)\,,\label{eq:2.8}\nonumber
\\
\langle 0 | \bar u(z) i\gamma_5 d(-z) | \pi(P)\rangle & = &
\frac{f_\pi m_\pi^2}{m_u+m_d}\, \int_0^1 du \, e^{i\xi pz}\,
\phi_{p}(u)\,,
\label{eq:2.11}\nonumber
\\
\langle 0 | \bar u(z) \sigma_{\alpha\beta}\gamma_5 d(-z) |
\pi(P)\rangle & = &-\frac{i}{3}\, \frac{f_\pi
  m_\pi^2}{m_u+m_d}  (p_\alpha z_\beta-
p_\beta z_\alpha) \int_0^1 du \, e^{i\xi pz}\,\phi_{\sigma}(u)\,,
\label{eq:2.12}
\end{eqnarray}
where $\xi\equiv2u-1$, $\phi_\pi$ is the leading twist-2 distribution amplitude, $\phi_{(p,\sigma)}$ are of twist-3.
All the above distribution amplitudes $\phi=\{\phi_\pi,\phi_p,\phi_\sigma,g_\pi\}$
are normalized to unity: $\int_0^1 du\, \phi(u) = 1$.

There is one 3-particle distribution amplitudes of twist-3, defined as \cite{pilcda}
\begin{eqnarray}
\langle 0 | \bar u(z) \sigma_{\mu\nu}\gamma_5
  g_sG_{\alpha\beta}(vz) d(-z)| \pi^-(P)\rangle
& = & i\,\frac{f_\pi m_\pi^2}{m_u+m_d} \left(p_\alpha p_\mu
  g_{\nu\beta}^\perp - p_\alpha p_\nu
  g_{\mu\beta}^\perp - p_\beta p_\mu g_{\nu\alpha}^\perp + p_\beta
  p_\nu g_{\alpha\mu}^\perp \right) {\cal T}(v,pz)\,,\label{eq:3pT3}
\end{eqnarray}
where we used the following notation for the integral
defining the 3-particle distribution amplitude:
\begin{equation}
{\cal T}(v,pz) = \int {\cal D}\underline{\alpha} \, e^{-ipz(\alpha_u
  -\alpha_d + v\alpha_g)} {\cal T}(\alpha_d,\alpha_u,\alpha_g)\,.
\end{equation}
Here $\underline{\alpha}$ is the set of three momentum fractions
$\alpha_d$, $\alpha_u$, and $\alpha_g$. The integration measure is
\begin{equation}
\int {\cal D}\underline{\alpha} = \int_0^1 d\alpha_d d\alpha_u
d\alpha_g \delta(1-\alpha_u-\alpha_d-\alpha_g)\,.
\end{equation}
The 3-particle distribution amplitudes of twist-4 are
\begin{eqnarray}
\langle 0 | \bar u(z)\gamma_\mu\gamma_5
g_sG_{\alpha\beta}(vz)d(-z)|\pi^-(P)\rangle
& = & p_\mu (p_\alpha z_\beta - p_\beta z_\alpha)\, \frac{1}{pz}\, f_\pi
m_\pi^2 {\cal A}_\parallel(v,pz) + (p_\beta g_{\alpha\mu}^\perp -
p_\alpha g_{\beta\mu}^\perp) f_\pi m_\pi^2 {\cal A}_\perp(v,pz)\,,\hspace*{1cm}\nonumber\\
\langle 0 | \bar u(z)\gamma_\mu i
g_s\widetilde{G}_{\alpha\beta}(vz)d(-z)|\pi^-(P)\rangle\
& = & p_\mu (p_\alpha z_\beta - p_\beta z_\alpha)\, \frac{1}{pz}\, f_\pi
m_\pi^2 {\cal V}_\parallel(v,pz) + (p_\beta g_{\alpha\mu}^\perp -
p_\alpha g_{\beta\mu}^\perp) f_\pi m_\pi^2 {\cal V}_\perp(v,pz)\,,\hspace*{1cm}
\end{eqnarray}
where $\widetilde{G}_{\alpha\beta}$ is the dual field
$\widetilde{G}_{\alpha\beta}\equiv \frac{1}{2}\varepsilon_{\alpha\beta\gamma\delta}G^{\gamma\delta}$.

We also use the distribution amplitude given in Ref. \cite{pilcda}:
\begin{eqnarray}
\phi_\pi(u)    &=& 6 u (1-u) \left( 1 + a_2 C_2^{3/2}(\xi)\right)\,,\\
g_\pi(u)       &=& 1+(1 + {18\over7}a2 + 60\eta_3 + {20\over3}\eta_4)C_2^{1/2}(\xi)
                   + (-{9\over28}a2 - 6\eta_3\omega_3)C_4^{1/2}(\xi)\,,\\
{\mathbb A}(u) &=& 6u\bar u \left\{ \frac{16}{15} + \frac{24}{35} \,
  a_2 + 20 \eta_3 + \frac{20}{9} \,\eta_4 \right.\nonumber\\
& &\left. + \left( -\frac{1}{15} + \frac{1}{16}\, - \frac{7}{27}\, \eta_3
  \omega_3 - \frac{10}{27}\, \eta_4 \right) C_2^{3/2}(\xi) + \left
  ( -\frac{11}{210} \, a_2 - \frac{4}{135}\, \eta_3\omega_3 \right)
  C_4^{3/2}(\xi) \right\} \nonumber\\
& & {}+ \left(-\frac{18}{5}\, a_2 + 21\eta_4\omega_4 \right) \left\{ 2
  u^3 (10-15 u + 6 u^2)\ln u + 2\bar u^3 (10-15\bar u + 6 \bar u^2)
  \ln\bar u + u \bar u (2 + 13u\bar u)\right\}\,,\\
\mathbb{B}(u) &=& g_\pi(u)-\phi_\pi(u)\,,\\
\phi_p(u)     &=& 1 + \left(30\eta_3 -\frac{5}{2}\, \rho_\pi^2\right)
C_2^{1/2}(\xi) + \left(- 3 \eta_3 \omega_3-\frac{27}{20}\, \rho_\pi^2
  - \frac{81}{10}\, \rho_\pi^2 a_2\right)  C_4^{1/2}(\xi)\,,\\
\phi_\sigma(u)&=& 6u(1-u) \left\{1 + \left(5\eta_3 -\frac{1}{2}\,\eta_3\omega_3 - \frac{7}{20}\,
                  \rho_\pi^2 - \frac{3}{5}\,\rho_\pi^2 a_2 \right) C_2^{3/2}(\xi)\right\}\,,\\
{\cal T}(\underline{\alpha}) &=& 360\eta_3 \alpha_u\alpha_d\alpha_g^2
\left\{1+\omega_3\, \frac{1}{2}\left( 7\alpha_g-3\right)\right\}\,,\\
{\cal V}_\parallel(\underline{\alpha}) & = & 120
\alpha_u\alpha_d\alpha_g ( v_{00} + v_{10} (3\alpha_g-1)),\nonumber\\
{\cal A}_\parallel(\underline{\alpha}) & = & 120
\alpha_u\alpha_d\alpha_g a_{10} (\alpha_d-\alpha_u)\,,\\
{\cal V}_\perp(\underline{\alpha}) & = & -30 \alpha_g^2\left[ h_{00}(1-\alpha_g)
                                         +h_{01}\Big[\alpha_g(1-\alpha_g)-6\alpha_u\alpha_d\Big]
                                         +h_{10}\Big[\alpha_g(1-\alpha_g)-\frac{3}{2}(\alpha_u^2
                                         +\alpha_d^2)\Big]\right]\,,\\
{\cal A}_\perp(\underline{\alpha}) & = & 30 \alpha_g^2(\alpha_u-\alpha_d)\left[ h_{00}+h_{01}\alpha_g
                                            +\frac{1}{2}\,h_{10}(5\alpha_g-3)\right]\,,
\end{eqnarray}
where $C_n^m(\xi)$ are Gegenbauer polynomials.

The definitions and the specific forms  of the $\eta$ light-cone distribution amplitudes adopted in the text are similar to those of the pion.
For more details see Ref. \cite{pilcda}.


\end{document}